\documentclass[aps,prd,groupedaddress,nofootinbib,superscriptaddress,preprint]{revtex4}
\usepackage{amssymb}
\usepackage{graphicx}
\usepackage{amsmath,color}
\usepackage{epsfig}
\usepackage{hyperref}
\begin{document}
\title{Non-standard antineutrino interactions at Daya Bay}
\author{Rupert Leitner}
\email{Rupert.Leitner@cern.ch}
\affiliation{Institute of Particle and Nuclear Physics, Faculty of
Mathematics and Physics, Charles University in Prague, V
Hole\v{s}ovi\v{c}k\'ach 2, 180 00 Praha 8, Czech Republic}

\author{Michal Malinsk\'y}
\email{malinsky@ific.uv.es}
\affiliation{AHEP Group, Instituto de F\'{\i}sica Corpuscular,
  C.S.I.C. -- Universitat de Val{\`e}ncia \\
  Edificio de Institutos de Paterna, Apartado 22085,
  E--46071 Val{\`e}ncia, Spain }

\author{Bed\v{r}ich Roskovec}
\email{roskovec@ipnp.troja.mff.cuni.cz} \affiliation{Institute of Particle
and Nuclear Physics, Faculty of Mathematics and Physics, Charles
University in Prague, V Hole\v{s}ovi\v{c}k\'ach 2, 180 00 Praha 8,
Czech Republic}

\author{He Zhang}
\email{he.zhang@mpi-hd.mpg.de} \affiliation{Max-Planck-Institut
f{\"u}r Kernphysik, Postfach 103980, 69029 Heidelberg, Germany}

\begin{abstract}

We study the prospects of pinning down the effects of non-standard
antineutrino interactions in the source and in the detector at the
Daya Bay neutrino facility. It is well known that if the non-standard interactions in the detection
process are of the same type as those in the production, their net effect can be subsumed into a mere shift in the
measured value of the leptonic mixing angle
$\theta_{13}$. Relaxing this assumption, the ratio of the
antineutrino spectra measured by the Daya Bay far and near detectors
is distorted in a characteristic way, and good fits based on the
standard oscillation hypothesis are no longer viable. We show that,
under certain conditions,  three years of Daya Bay running can be sufficient to provide a clear hint of non-standard neutrino physics.
\end{abstract}
\maketitle

\section{Introduction}
\label{Sec:Introduction}

The past and ongoing neutrino oscillation experiments provide a firm
evidence that the neutrino flavour is changing throughout the
neutrino propagation. Except for some recent signals reported by
MINOS~\cite{Adamson:2011fa} and
MiniBooNE~\cite{AguilarArevalo:2010wv} (and also previously by
LSND~\cite{Aguilar:2001ty}) the vast majority of the data is
consistent with the hypothesis of neutrino flavour oscillations
driven by a pair of mass-squared differences: $\Delta m_{31}^{2}=2.45\pm 0.09
 \times 10^{-3}~{\rm eV}^2$ 
 (or $\Delta m_{31}^{2}=-(2.34^{+0.10}_{-0.09})
 \times 10^{-3}~{\rm eV}^2$ if an inverse neutrino mass hierarchy is realized), often called the atmospheric mass-squared difference, and the solar mass-squared
difference $\Delta m_{21}^{2}=7.59^{+0.20}_{-0.18} \times 10^{-5}~{\rm
eV}^2$, together with a pair of the corresponding mixing angles: $\sin^2\theta_{23}=0.51\pm 0.06$ (or  $\sin^2\theta_{23}=0.52\pm 0.06$ the inverse hierarchy case) and
$\sin^2\theta_{12}=0.312^{+0.017}_{-0.015}$ ~\cite{Schwetz:2011qt}. This,
however, requires at least two of the oscillating neutrinos to be
massive.

By construction, neutrinos are massless in the Standard Model (SM).
Thus, a lot of  effort has been spent on devising SM extensions
that could not only accommodate the unprecedented smallness of the
light neutrino mass scale and all the peculiarities of the leptonic
mixing pattern, but also provide specific new physics signals, thus admitting for a further experimental scrutiny. In some cases, such new
physics effects could even be expected to be within the reach of
near future experimental facilities. In this respect, the seesaw
approach
\cite{Minkowski:1977sc,Yanagida:1979as,GellMann:1980vs,Mohapatra:1979ia,Schechter:1980gr,Lazarides:1980nt,Mohapatra:1980yp,Foot:1988aq},
in which the smallness of the absolute neutrino mass scale is usually
 linked to  a very specific type of high energy dynamics, represents a particularly plausible
model-building paradigm.

Each  dynamical realization of the seesaw picture makes some
kind of new physics effects appear, at least at a certain level. This, in
turn, makes the neutrino sector an ideal probe to physics beyond the
SM. For instance, the Majorana nature of the light neutrinos
inherent to the seesaw framework provides characteristic
lepton-number-violating signals at low energies like, e.g.,
the neutrinoless double beta decay, or, if kinematically accessible,
same-sign di-lepton production at colliders, see, e.g., \cite{Han:2006ip}. Similarly, besides
neutrino oscillations, the flavour structure of the lepton sector can
be tested in lepton-flavour-violating processes such as $\mu\to
e\gamma$ or, for example, trilepton collider events, c.f., \cite{delAguila:2008cj}.

A full exploration of such new physics signals generally requires a
very good knowledge of the {leptonic flavour} mixing angles governing
the neutrino oscillation phenomena. In particular, the smallest
mixing angle $\theta_{13}$, which plays a central role in the
leptonic CP violation, is still to be determined (with a current $90\%$~C.L.
upper bound of $\sin^{2} 2\theta_{13}\lesssim 0.17$ 
reported by the CHOOZ collaboration~\cite{Apollonio:2002gd} and first indications of its non-zero value obtained by T2K \cite{T2K} and MINOS \cite{MINOS} experiments).

The Daya Bay neutrino oscillation experiment \cite{Guo:2007ug} is
designed to perform a precision determination of $\theta_{13}$ with
a potential to improve the CHOOZ limit by one order of magnitude.
More specifically, the $\sin^{2} 2\theta_{13}$-sensitivity of Daya Bay is anticipated to reach 0.01 at $90\%$ confidence level over the entire allowed range of $\Delta m_{31}^{2}$, see Fig. 3.12. in \cite{Guo:2007ug}. At the assumed best fit point $\Delta m_{31}^{2}=2.51 \times 10^{-3}{\rm eV}^{2}$ the expected sensitivity is around 0.008 with 3 years of data.
Such a highly ambitious goal relies on a very good control over the systematics,
which is achieved by employing a unique set of eight identical detectors deployed at three different
locations optimized for monitoring the antineutrino rates from the
six reactors. A similar experimental setup is also adopted by the upcoming
Double Chooz experiment~\cite{Ardellier:2006mn} and RENO \cite{Kim:2010zzd} experiments.

In combination with the large statistics due to the huge flux of
antineutrinos produced in the nearby nuclear reactors, the unprecedented
accuracy of this new generation of reactor experiments can make them
sensitive to the new physics effects, at least at a certain level.
For instance, if the new physics sector couples to hadrons and the
relevant scale is not very high, one can expect non-standard
 interactions (NSI's) in the antineutrino production and
detection processes as well as non-standard matter effects the antineutrinos
experience throughout the propagation process. Similarly,  new neutral fermions can mix with
the three SM active neutrinos, which would result in an effective non-unitarity
of the leptonic mixing matrix entering the relevant oscillations
probabilities.

The NSI's in reactor neutrino experiments have been discussed
previously in, e.g., ~\cite{Kopp:2007ne,Ohlsson:2008gx}, especially
when the production and detection processes (and the corresponding non-standard effects) are assumed to be just inverse of each other. In particular, it has been shown that in such a
case the NSI effects can be subsumed into a mere shift in the
measured value of the effective mixing angle~$\theta_{13}$.

In this work, we study the NSI's in reactor antineutrino
experiments in a general case when the assumption that the
source and at the detector processes including the non-standard effects are
just inverse of each other is dropped. This, in turn, leads to a specific
distortion in the ratio of the antineutrino spectra measured in the
far and in the near detectors which can not be entirely transformed
away by mere shifts in the relevant oscillation parameters, i.e., $\theta_{13}$ and $\Delta m_{31}^{2}$. We show that,
under certain conditions,  three years of Daya Bay running can be sufficient to provide a clear hint of non-standard neutrino physics. 

The remainder of the manuscript is organized as follows: In
Sec.~\ref{sec:nsiformula}, we present the general formalism and
derive the relevant antineutrino survival probability formulas used in the
subsequent analysis. Two basic scenarios corresponding to
qualitatively different shapes of NSI's are specified in
Sect.~\ref{sec:settings}, and a detailed analysis of the
observability of such effects at Daya Bay is performed in
Sect.~\ref{sec:results}. Finally, we conclude in
Sect.~\ref{sect:summary}.

\section{Non-standard  interactions in reactor antineutrino oscillations}
\label{sec:nsiformula}

\subsection{Non-standard interactions in the antineutrino sources and detectors}
\label{sect:settings}
In what follows, we adopt the standard SPD (source, propagation, and detector)
approach~\cite{Huber:2002bi} to consider the antineutrino
oscillation process in a reactor antineutrino experiment. In the
presence of NSI's, the antineutrino states produced in the source 
as well as those observed in the detector can be treated as superpositions of
pure orthonormal flavour states
\begin{equation}\label{eq:nus}
|\overline{\nu^{s}_\alpha} \rangle  =   |\overline{\nu_\alpha} \rangle +
\sum_{\beta=e,\mu,\tau} \varepsilon^{s*}_{\alpha\beta}
|\overline{\nu_\beta}\rangle   \ , \qquad
\langle \overline{\nu^d_\beta}|  =   \langle
 \overline{\nu_\beta} | + \sum_{\alpha=e,\mu,\tau}
\varepsilon^{d*}_{\alpha \beta} \langle  \overline{\nu_\alpha}  |   \ ,
\end{equation}
where the superscripts `$s$' and `$d$' denote the source and the
detector, respectively. Note that there is no need to include the
appropriate normalization factors in expressions (\ref{eq:nus})
because  we are going to be interested only in ratios of the survival
probabilities in the near and far detectors where such factors cancel. 

The current experimental bounds on the NSI parameters mainly come
from the lepton flavour violating decays $\ell_\alpha \to \ell_\beta
\gamma$, the universality test of weak interactions and the
invisible decay width of the $Z$-boson. Model-independent studies
indicate that the upper limits on the NSI parameters $\varepsilon^{s,d}_{\alpha,\beta}$ are typically
in the ballpark of $10^{-1}$ to $10^{-2}$, see e.g.~\cite{Biggio:2009nt} and references therein.

{
To keep the analysis as general as possible, in what follows we shall consider two basic physical situations, namely, $\varepsilon_{s}= \varepsilon_{d}^{\dagger}$ (to be called case I, cf. Section \ref{sect:CaseI}) and also $\varepsilon_{s}\neq \varepsilon_{d}^{\dagger}$ (case II, see Section \ref{sect:inequivalent}) and provide a detailed performance analysis of the Daya Bay experimental setting in each case. 

Few comments are worth here: First, it is not very common to assume $\varepsilon_{s}\neq \varepsilon_{d}^{\dagger}$ as, in such a case, given the inverse microscopic nature of the relevant source and detection processes in standard reactor neutrino experiments, one usually concludes that these two quantities should be equal, see e.g. \cite{Kopp:2007ne}. This, however, is based on several implicit assumptions, in particular: 1) exact CPT (and Lorentz) invariance and 2) factorizability of the source, propagation and detection processes.  As for the former, CPT violation (implying also the Lorentz invariance breakdown), for instance, makes it possible to have masses and mixings different for neutrinos and antineutrinos. Hence, the mixing matrix entering the detection process can be different from the one governing the antineutrino production and, thus, the standard oscillation formula is not valid even if all the NSI parameters are set to zero. Nevertheless, at least to the leading order in $\varepsilon_{s}$ and $\varepsilon_{d}$\,, such an effect can be still accounted for within the standard SPD formalism by a suitable redefinition of the NSI parameters, thus generating an effective departure from the assumed $\varepsilon_{s}=\varepsilon_{d}^{\dagger}$ case, cf. formula (\ref{eq:J}) in Section \ref{NSIformulae}. An interested reader can find further comments in, e.g., the recent work \cite{Antusch:2008zj} and references therein. Needless to say, these considerations became especially relevant with the recent claims of observation of possibly superluminal neutrinos in the OPERA experiment~\cite{opera:2011zb}. Concerning 2), it is clear that as long as the antineutrino is produced in the source by the classical SM charged current interaction no departure from the basic case I setting can be expected because all the flavour-blind standard nuclear effects such as, e.g., the enhancement of scalar and/or tensor modes etc., simply factorize out and, hence, contribute only to the change of normalization of the total neutrino fluxes. In this respect, the $\varepsilon_{s}\neq \varepsilon_{d}^{\dagger}$ case can be viewed as a model-independent parametrization of non-factorizable flavour-dependent new physics effects which, at low energies, can correspond to higher-dimensional terms in the effective lagrangian such as, e.g., $(N\overline{N})^{2}\nu\overline{\nu}$ (which can be relevant also to other processes such as neutron star cooling, see, e.g., \cite{Stoica:2004zy}) etc.  Therefore, though possibly marginal from the perspective of the conventional extensions of the SM, we still find this setting worth a closer look, the more that Daya Bay can perform very well in this case, see Sections  \ref{num:caseIIa} and \ref{num:caseIIb}.  
}

\subsection{Non-standard interactions in the antineutrino propagation}

The propagation of antineutrino flavour eigenstates from the sources to
the detectors is governed by the effective Hamiltonian
\begin{eqnarray}\label{eq:Hamiltonian}
\hat{H} & = &  H_0 + H_m + H_{\rm NSI} = \frac{1}{2E}U^{*} {\rm
diag} (m^2_1,m^2_2,m^2_3) U^T- {\rm diag} (V_{\rm CC},0,0) - V_{\rm
CC} \varepsilon^{m*} \, ,
\end{eqnarray}
where $\varepsilon^m$
is a Hermitian matrix parametrizing the NSI's throughout the antineutrino
propagation and $V_{\rm CC}=\sqrt{2}G_F N_e$ arises due to effects of the coherent forward
scattering in matter (with $N_e$ denoting the electron number density along the
antineutrino trajectory). Barring the irrelevant Majorana phases, the vacuum leptonic mixing matrix
$U$ is conveniently parametrized by three
mixing angles $\theta_{12}$, $\theta_{23}$ and $\theta_{13}$ and one Dirac CP phase $\delta$ \cite{Amsler:2008zz}
\begin{eqnarray}\label{eq:parametrization}
U & = & \left(
\begin{matrix}c^{}_{12} c^{}_{13} & s^{}_{12} c^{}_{13} & s^{}_{13}
{\rm e}^{-{\rm i}\delta^{}} \cr -s^{}_{12} c^{}_{23}-c^{}_{12}
s^{}_{23} s^{}_{13} {\rm e}^{{\rm i} \delta^{}} & c^{}_{12}
c^{}_{23}-s^{}_{12} s^{}_{23} s^{}_{13} {\rm e}^{{\rm i} \delta^{}}
& s^{}_{23} c^{}_{13} \cr
 s^{}_{12} s^{}_{23}-c^{}_{12} c^{}_{23} s^{}_{13}
{\rm e}^{{\rm i} \delta^{}} & -c^{}_{12} s^{}_{23}-s^{}_{12}
c^{}_{23} s^{}_{13} {\rm e}^{{\rm i} \delta^{}} & c^{}_{23}
c^{}_{13}\end{matrix} \right) \ ,
\end{eqnarray}
with $c^{}_{ij} \equiv \cos \theta^{}_{ij}$ and $s^{}_{ij} \equiv
\sin \theta^{}_{ij}$ (for $ij=12$, $13$ and $23$).
The full effective Hamiltonian \eqref{eq:Hamiltonian} is
diagonalized via a unitary transformation
\begin{eqnarray}\label{eq:effH}
\hat{H} = \frac{1}{2E} \hat{U}^{*} {\rm diag}\left( \hat{m}^2_1,
\hat{m}^2_2, \hat{m}^2_3 \right) \hat{U}^{T} \, ,
\end{eqnarray}
where $\hat{m}_i$ ($i=1,2,3$) denote the effective masses of
neutrinos and $\hat U$ is the effective leptonic
mixing matrix in matter.

The size of the matter effect driven by the $\sqrt{2} G_{F} N_{e}$
term amounts to around $1.1 \times 10^{-7}$~eV$^{2}$/MeV for Earth
crust with density of 2.8 g/cm$^{3}$. Even for the highest values of
reactor antineutrino energies of around 10 MeV, this number is
about 40 times smaller than the value of $\Delta m^{2}_{21}
/(2E)=3.8\times 10^{-6}$~eV$^{2}$/MeV and about 1100 times smaller
than $\Delta m^{2}_{32} /(2 E)=1.2\times 10^{-4}$~eV$^{2}$/MeV. This
indicates that Earth matter effects are very small and can be
safely neglected. Hence, we take $\hat H \simeq H_0$ or,
equivalently, set $V_{\rm CC}=0$ in Eq.~\eqref{eq:Hamiltonian}.

\subsection{The antineutrino survival probability}
\label{NSIformulae}
With the NSI effects at play, the electron
antineutrino survival probability amplitude  ${\cal A}(\overline{\nu^s_e} \rightarrow \overline{\nu^d_e};L)\equiv {\cal A}_{ee}(L)$ is given
by
\begin{equation}\label{eq:effA1}
{\cal A}_{ee}(L)  =   \langle \overline{\nu^d_e} | {\rm e}^{-{\rm
i} {H} L} |\overline{\nu^s_e} \rangle = ({1} +
{\varepsilon^{d*}})_{\rho e} A_{\gamma\rho}\left({1} +
{\varepsilon^{s*}}\right)_{e\gamma} = \left[ { A} + {\varepsilon^{s*}} { A} + { A} {\varepsilon^{d*}} +
{\varepsilon^{s*}} { A} {\varepsilon^{d*}} \right]_{ee} \,,
\end{equation}
where $L$ is the propagation distance and $A$
is a coherent sum over the contributions of all the mass eigenstates
$\nu_i$
\begin{eqnarray}\label{eq:A}
{A}_{\alpha\beta} = \sum_i  U_{\alpha i} U^{*}_{\beta i} {\rm
e}^{-{\rm i} \frac{m^2_i L}{2E}}\ .
\end{eqnarray}
The antineutrino survival probability is then given by
$
P(\overline{\nu^s_e} \rightarrow \overline{\nu^d_e}) = \left| {\cal A}_{{ee}}(L)
\right|^2
$.
For completeness, let us remark that a corresponding neutrino oscillation amplitude
can be readily obtained from (\ref{eq:effA1}) with a substitution $(U^{*},\varepsilon^{*})\to(U,\varepsilon)$. It should also be stressed
that only the first row of $\varepsilon^s$ and the first column of
$\varepsilon^d$ are relevant to the $ee$-type transition amplitude.
Namely, the NSI parameters $\varepsilon^s$ and $\varepsilon^d$
involved in reactor neutrino experiment contain at least one flavour
index~$e$.

Inserting formula (\ref{eq:A}) into Eq.~\eqref{eq:effA1} one
arrives at the full antineutrino oscillation probability
\begin{equation}\label{eq:P1}
P(\overline{\nu^s_e} \rightarrow \overline{\nu^d_e}) = \sum_{i,j}
{\cal J}^i  {\cal J}^{j*}  - 4 \sum_{i>j} {\rm Re} ({\cal J}^i {\cal
J}^{j*}  )\sin^2\left(\frac{\Delta   m^2_{ij}L}{4E}\right) + 2 \sum_{i>j}{\rm Im}
( {\cal J}^i  {\cal J}^{j*}  ) \sin\left(\frac{ \Delta   m^2_{ij} L}{2 E}\right)
\, ,
\end{equation}
where $\Delta m^2_{ij}=m^2_i-m^2_j$, and
\begin{eqnarray}\label{eq:J}
{\cal J}^i =  U_{e i}  U^*_{e i} + \sum_\gamma \varepsilon^{s*}_{e
\gamma}  U_{\gamma i} U^*_{e i}+ \sum_\gamma
\varepsilon^{d*}_{\gamma e} U_{e i} U^*_{\gamma i} +
\sum_{\gamma,\rho} \varepsilon^{s*}_{e\gamma} \varepsilon^{d*}_{\rho
e} U_{\gamma i} U^*_{\rho i}  \, .
\end{eqnarray}
In the $\varepsilon^{s,d} \to 0$ limit, Eq.~\eqref{eq:P1} reduces to the
standard survival probability.

In this study, the quantity of our main interest  is the third term in Eq.~\eqref{eq:P1} which, being linear in the sine of  $L/E$, does not play any role in the standard oscillation case. In this respect,  a potential deviation from the ``standard'' quadratic-sine $L/E$ dependence in an oscillation experiment can be interpreted as a hint of  non-standard antineutrino interactions,  in particular if such an anomaly exhibits the characteristic linear-sine $L/E$ shape. 

\subsection{Series expansion of the antineutrino survival probability}

In practice, given the finite precision of the experimental inputs, it is very convenient to expand the survival probability (\ref{eq:P1}) around the standard oscillation formula in terms of the relevant small
parameters, in particular $\varepsilon^{s,d}$ which are all expected to be at most at the few per-cent level, c.f.~\cite{Biggio:2009nt} and references therein. In addition, $\theta_{13}$ is small compared to the
other mixing angles (with the current CHOOZ upper limit of $\sin^{2}
2\theta_{13}\lesssim 0.17$) and, hence, it amounts to another useful
expansion parameter. Moreover, for the Daya Bay far detector, also the
oscillation term $\Delta m^2_{21} L /(2 E)$ turns out to be at the
level of $10^{-1}$ to $10^{-2}$ and, as such, it can also be viewed
as a small quantity.

Taking all this into account, we obtain the following expanded form
of the relevant  electron antineutrino survival probability
\begin{eqnarray}\label{eq:Pexpand}
P(\overline{\nu_{e}^{s}}\to \overline{\nu_{e}^{d}}) &\simeq&
P(\overline{\nu_{e}}\to \overline{\nu_{e}})_{\rm SM} -4\left[{\rm
Re}\left( \varepsilon _{e\mu }^{s}e^{-{\rm i}\delta }+\varepsilon _{\mu
e}^{d}e^{{\rm i}\delta }\right) s_{23}s_{13}+{\rm Re}\left( \varepsilon
_{e\tau }^{s}e^{-{\rm i}\delta }+\varepsilon _{\tau e}^{d}e^{{\rm i}\delta
}\right) c_{23}s_{13} \right.\nonumber \\ &+&\left. {\rm Re}\left(
\varepsilon _{e\mu }^{s}\varepsilon _{\mu e}^{d}\right)
s_{23}^{2}+{\rm Re}\left( \varepsilon _{e\tau }^{s}\varepsilon
_{\tau e}^{d}\right) c_{23}^{2}+{\rm Re}\left( \varepsilon _{e\mu
}^{s}\varepsilon _{\tau e}^{d}+\varepsilon _{e\tau }^{s}\varepsilon
_{\mu e}^{d}\right) s_{23}c_{23}\right]\sin ^{2}\left( \frac{\Delta m_{32}^{2}L}{4E}\right) \nonumber  \\
&+&2\left[ {\rm Im}\left( \varepsilon _{e\mu }^{s}e^{-{\rm i}\delta
}+\varepsilon _{\mu e}^{d}e^{{\rm i}\delta }\right) s_{23}s_{13}+{\rm
Im}\left( \varepsilon _{e\tau }^{s}e^{-{\rm i}\delta }+\varepsilon _{\tau
e}^{d}e^{{\rm i}\delta }\right) c_{23}s_{13} \right. \nonumber \\ &+&
\left. {\rm Im}\left( \varepsilon _{e\mu }^{s}\varepsilon _{\mu
e}^{d}\right) s_{23}^{2}+{\rm Im}\left( \varepsilon _{e\tau
}^{s}\varepsilon _{\tau e}^{d}\right) c_{23}^{2}+{\rm Im}\left(
\varepsilon _{e\mu }^{s}\varepsilon _{\tau e}^{d}+\varepsilon
_{e\tau }^{s}\varepsilon
_{\mu e}^{d}\right) s_{23}c_{23}\right] \sin \left( \frac{\Delta m_{32}^{2}L}{2E}\right) \nonumber  \\
&+&2\left[ {\rm Im}\left( \varepsilon _{e\mu }^{s}+\varepsilon _{\mu
e}^{d}\right) c_{12}s_{12}c_{23}-{\rm Im}\left( \varepsilon _{e\tau
}^{s}+\varepsilon _{\tau e}^{d}\right) c_{12}s_{12}s_{23}\right]
\left(
\frac{\Delta m_{21}^{2}L}{2E}\right) \\
&+& {\cal O}\left[\varepsilon^3, s^3_{13}, \varepsilon^2 s_{13},
\varepsilon s^2_{13},\varepsilon s_{13} \left( \frac{\Delta
m_{21}^{2}L}{2E} \right) , \varepsilon \left( \frac{\Delta
m_{21}^{2}L}{2E} \right)^2,s_{13}^{2} \left( \frac{\Delta
m_{21}^{2}L}{2E} \right)\right]\, ,\nonumber
\end{eqnarray}
where $P(\overline{\nu_{e}}\to \overline{\nu_{e}})_{\rm SM}$
corresponds to the standard oscillation probability, i.e.,
the one without NSI's which is approximately given by
\begin{equation}\label{eq:PSM}
P(\overline{\nu_{e}}\to \overline{\nu_{e}})_{\rm SM}\simeq 1 - 4
s^2_{13} \sin^2\left(\frac{\Delta m^2_{32}L}{4E}\right) - 4 s^2_{12}
c^2_{12}\left(\frac{\Delta m^2_{21}L}{4E}\right)^{2} + {\cal
O}\left[s^3_{13}, s_{13}^{2} \left( \frac{\Delta m_{21}^{2}L}{2E}
\right)\right].
\end{equation}
Inspecting Eq.~(\ref{eq:Pexpand}) one can recognize three
qualitatively different non-standard contributions to
$P(\overline{\nu_{e}^{s}}\to \overline{\nu_{e}^{d}})$: In the first
two lines there is a CP-even term quadratic in sine of ${\Delta
m_{32}^{2}L}/(4E)$ which, as expected, may affect the
determination of the mixing angle $\theta_{13}$. The remaining three
lines denote the CP-odd NSI effects corresponding to two different
kinematical regimes characterized by $\Delta m_{32}^{2}$ and $\Delta
m_{21}^{2}$, respectively. Notice that in the standard
parametrization \eqref{eq:parametrization}, the Dirac
CP-violating phase groups only with the former factor. It is also worth noticing that the term proportional to $\Delta
m_{21}^{2}$ tends to be further suppressed in the ``flavour-blind'' setting (with
$\varepsilon_{e\mu}^{s,d}=\varepsilon^{s,d}_{e\tau}$) because of the
proximity of $s_{23}$ and $c_{23}$.

It shall be noted that both the standard and the NSI transition
probabilities depend on the neutrino mass hierarchy. For the standard
oscillations, the hierarchy-sensitive terms are of the order of
$s^{2}_{13}{\Delta m^{2}_{21} L}/({2 E})$ and thus can be
consistently neglected in Eq.~(\ref{eq:PSM}). The NSI-dependent
terms in Eq.~\eqref{eq:Pexpand}, however, contain a term linear in
sine of ${\Delta m_{32}^{2}L}/({2E})$ which, indeed, differs in sign
in the normal and in the inverted hierarchy schemes, respectively. Since, however, we do not expect any distinctive NSI
features to be large enough to discriminate among these two settings
(although they would certainly differ in details), in what follows,
we shall deliberately stick to the normal hierarchy case, i.e.,
assume $\Delta m^{2}_{32} > 0$.

\subsection{Notation and conventions}

In what follows we shall adopt the following parametrization:
\begin{equation}
\label{eq:notation1}
\varepsilon^s_{e\alpha} \equiv \left|\varepsilon_{\alpha}^{s}\right| e^{{\rm i}\phi_{\alpha}^{s}}
\,, \qquad
\varepsilon^d_{\alpha e} \equiv \left|\varepsilon_{\alpha}^{d}\right| e^{-{\rm i}\phi_{\alpha}^{d}}\,,
\end{equation}
where the universal $e$ index was dropped for simplicity. It is also
convenient to define the source and detector phase averages
$\Phi_{\alpha}$ and differences $\Delta\phi_{\alpha}$, respectively:
\begin{equation}
\label{notation2}
\Phi_{\alpha}\equiv \tfrac{1}{2}(\phi_{\alpha}^{d}+\phi_{\alpha}^{s})\,, \qquad
\Delta\phi_{\alpha} \equiv  \tfrac{1}{2}(\phi_{\alpha}^{d}-\phi_{\alpha}^{s})\,.
\end{equation}
The latter has a clear physical meaning: indeed, for all
$\Delta\phi_{\alpha}\to 0$ (together with
$|\varepsilon^{s}_{\alpha}|\to |\varepsilon^{d}_{\alpha}|$) one
recovers a limit in which the non-standard antineutrino interactions in the detection process are of the same kind as those in the production.

\section{Specific settings\label{sec:settings}}

In what follows, we shall discuss two simple but phenomenologically
interesting shapes of NSI's and discuss the relevant effects in the
reactor antineutrino experiments.

\subsection{\label{sect:CaseI}Case I: $\varepsilon^s_{\alpha} = \varepsilon^{d*}_{\alpha}$}

We start with the simplest case characterized by the assumption
$\varepsilon^s_{\alpha} = \varepsilon^{d*}_{\alpha} \equiv
|\varepsilon_{\alpha}|e^{{\rm i}\phi_{\alpha}}$ which corresponds to the
situation where the production and the detection processes (including the associated non-standard interactions) are just inverse of each other.
The relevant antineutrino survival probability (\ref{eq:Pexpand}) is then reduced to
\begin{eqnarray}\label{eq:PcaseI}
P(\overline{\nu_{e}^{s}}\to \overline{\nu_{e}^{d}}) &\simeq&
P(\overline{\nu_{e}}\to \overline{\nu_{e}})_{\rm SM} -4\left\{
s_{23}^{2}|\varepsilon_{\mu}|^{2}+
c_{23}^{2}|\varepsilon_{\tau}|^{2}+2s_{23}c_{23} |\varepsilon_{\mu}|
|\varepsilon_{\tau}|\cos(\phi_{\mu}-\phi_{\tau})\right.
\nonumber\\
& +& 2 s_{13}\left.\left[
s_{23}|\varepsilon_{\mu}|\cos(\phi_{\mu}-\delta)+
c_{23}|\varepsilon_{\tau}|\cos\left(\phi_{\tau}-\delta\right)
\right]\right\} \sin^{2}\left(\frac{\Delta m_{32}^{2}L}{4E}\right)
\, .
\end{eqnarray}
Remarkably, the linear sine-dependent term in Eq.~\eqref{eq:Pexpand}
vanishes and the NSI effects enter the survival probability as a
mere global shift of the oscillation amplitude. This amounts to a
shift in the ``effective'' reactor mixing angle
\begin{eqnarray}\label{eq:th13eff}
s_{13}^{2}\;\to \;\tilde s^2_{13} & = & s^2_{13}+\left.
s_{23}^{2}|\varepsilon_{\mu}|^{2}+
c_{23}^{2}|\varepsilon_{\tau}|^{2}+2s_{23}c_{23}|\varepsilon_{\mu}|
|\varepsilon_{\tau}|\cos(\phi_{\mu}-\phi_{\tau})\right.
\nonumber\\
& +& 2 s_{13} \left[
s_{23}|\varepsilon_{\mu}|\cos(\phi_{\mu}-\delta)+
c_{23}|\varepsilon_{\tau}|\cos(\phi_{\tau}-\delta) \right] \, .
\end{eqnarray}
Namely, the oscillation probability is given by the standard
formula (\ref{eq:PSM}) with $\theta_{13}$ replaced by the effective
mixing angle $\tilde \theta_{13}$. Thus, there is no way to
discriminate such an NSI effect from standard oscillations in reactor
antineutrino experiments. It is also worth noting that the CP phase
differences enter Eq.~(\ref{eq:th13eff}) via cosines only which is,
indeed, justified by the CP properties of the survival probability
in the setting under consideration.

\begin{figure}[t]
\begin{center}\vspace{0.cm}
\includegraphics[width=9cm]{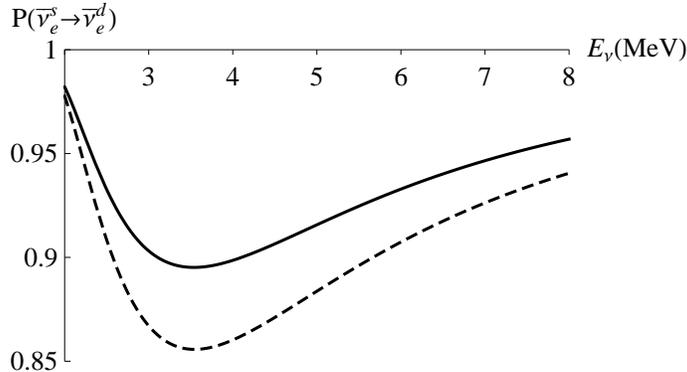}
\vspace{0.5cm} \caption{\label{fig:fig1} The oscillation probability for $\sin^{2} 2\theta_{13}=0.1$, $\delta=0$ and $L=1.8$~km with no NSI's (solid line) and with NSI's  corresponding to Case I in Sect.~\ref{sect:CaseI} with $\varepsilon^s_{\alpha} = \varepsilon^{d*}_{\alpha}$ (dashed line) where we adopted $\varepsilon_{\mu}= \varepsilon_{\tau}=0.02$ (with both $\phi_{\mu}$ and $\phi_{\tau}$  fixed to zero). For the other oscillation parameters, the
best-fit values have been assumed, c.f.
Ref.~\cite{Schwetz:2011qt}.} \vspace{-0.cm}
\end{center}
\end{figure}

In Fig.~\ref{fig:fig1}, we display the standard and the modified oscillation probability in the NSI presence as a
function of the antineutrino energy in a detector at the ``ideal''
distance $L=1.8$ km  (optimized for the highest count rate 
at $E\sim 4$~MeV) from the source.
The ``depth'' of the first  oscillation minimum (the solid line for the standard oscillations) changes significantly if the NSI effects are turned on (dashed line); however, the energy of the minimum determined by the
neutrino mass-squared differences remains essentially unchanged.\footnote{Let us
remark that the energies corresponding to the two relevant minima
are not exactly the same due to the presence of the sub-leading
terms proportional to $\left({\Delta m^2_{21}L}/{4E}\right)^{2}$ in
Eqs.~\eqref{eq:Pexpand} and \eqref{eq:PSM}.}

Nevertheless, though the reactor antineutrino experiments in this case cannot
distinguish the NSI's from a true mixing angle on their own, they can still
provide a useful piece of information in combination with other types of experiments such as, e.g., accelerator experiments, superbeams, beta-beams,
neutrino factories, etc. In particular, if these searches report 
different values of $\theta_{13}$, NSI's could be responsible for the mismatch.

\subsection{Case II: $ \varepsilon^s_{\alpha} \neq \varepsilon^{d*}_{\alpha}$\label{sect:inequivalent}}
{ As already mentioned in Section \ref{sect:settings}, we do not intend to confine ourselves entirely to the ``canonical'' case with $\varepsilon^s_{\alpha} = \varepsilon^{d*}_{\alpha}$, but  rather prefer to keep the mind open also to the intriguing $\varepsilon^s_{\alpha} \neq \varepsilon^{d*}_{\alpha}$ possibility. Indeed, the distortion of the shape of the detected antineutrino spectra with respect to the standard oscillation picture expected with such a choice of the NSI parameters in the effective quantum-mechanical SPD picture can mimic (at the leading order) a wide class of new physics effects such as, e.g., a net CPT violation or, for instance, non-factorizable beyond-Standard-Model effects in the production processes, cf. Section~\ref{sec:nsiformula}.}

\begin{figure}[t]
\begin{center}\vspace{0.cm}
\includegraphics[width=9cm]{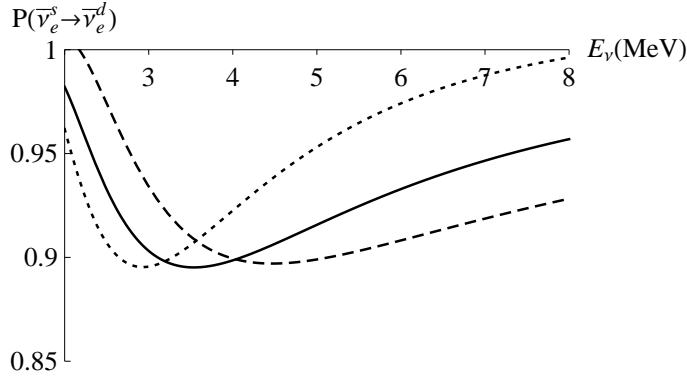}
\vspace{0.5cm} \caption{\label{fig:fig2} The theoretical oscillation
probability with no NSI (solid line) and with NSI's (dashed and
dotted lines) for $L=1.8$ km. For the sake of illustration, we adopt a flavour-universal scheme with all the relevant NSI parameters at the same level of $|\varepsilon_{\mu,\tau}|=0.04$ with $\Delta \phi_{\mu,\tau} =\frac{\pi}{2}$. The other neutrino mixing
parameters are the same as in Fig.~\ref{fig:fig1}. The dashed line corresponds to $\Phi_{\mu,\tau}= \frac{8}{5} \pi$ whereas the dotted one to $\Phi_{\mu,\tau}=\frac{2}{5} \pi$, respectively.} \vspace{-0.cm}
\end{center}
\end{figure}

As a consequence, the
terms linear in sine in formula (\ref{eq:Pexpand}) are exposed and
the relevant NSI effects can no longer be completely subsumed into a
shift of the effective mixing angle $\tilde\theta_{13}$. This,
besides the change of the ``depth'' of the first oscillation minimum
(c.f. Figure \ref{fig:fig1}), leads also to a shift in its energy,
as illustrated in Figure~\ref{fig:fig2}.
In particular, the dip can be shifted by as much as one ${\rm MeV}$ in both directions,
depending on the specific choice of the NSI parameters. 

In what follows we shall focus on two specific realizations of this
setting, namely, the case when the magnitude of the NSI parameters
differs between the production and detection processes (Case IIa)
and the case when the relevant NSI parameters are of the same size
but differ by their phases (Case IIb).  Both these cases are studied
numerically in Sect.~\ref{numerics}.

\subsubsection{\label{sect:caseIIa}Case IIa: Non-standard interactions in source only}

Let us exemplify the first option on a specific setting where the
NSI's exhibit themselves only in the production processes, i.e.,
taking $\varepsilon^{d}=0$. Given that, the general formula
(\ref{eq:Pexpand}) simplifies into
\begin{eqnarray}
P(\overline{\nu_{e}^{s}}\to \overline{\nu_{e}^{d}}) &\simeq&
P(\overline{\nu_{e}}\to \overline{\nu_{e}})_{\rm SM} \nonumber \\
&-&4s_{13}\left[ s_{23}|\varepsilon_{\mu}|\cos(\phi_{\mu}-\delta)+
c_{23}|\varepsilon_{\tau}|\cos(\phi_{\tau}-\delta) \right]
\sin^{2}\left(\frac{\Delta m_{32}^{2}L}{4E}\right)\nonumber\\
& +& 2 s_{13}\left[
s_{23}|\varepsilon_{\mu}|\sin(\phi_{\mu}-\delta)+
c_{23}|\varepsilon_{\tau}|\sin(\phi_{\tau}-\delta)
\right]
\sin\left(\frac{\Delta m_{32}^{2}L}{2E}\right)\nonumber\\
& +&  2 s_{12}c_{12}\left( c_{23}|\varepsilon_{\mu}|\sin\phi_{\mu}-
s_{23}|\varepsilon_{\tau}|\sin\phi_{\tau} \right)\left( \frac{\Delta
m_{21}^{2}L}{2E} \right) \, ,\label{general:sonly:raw}
\end{eqnarray}
where we used $\varepsilon_{\alpha}\equiv
\varepsilon^{s}_{\alpha}$ and $\phi_{\alpha}\equiv
\phi^{s}_{\alpha}$. It is worth noting that the genuine NSI effect
(due to the last two terms) is proportional to sines of differences
of the Dirac CP phase $\delta$ and the CP phases of the NSI
parameters $\phi_{\alpha}$, as expected for a CP-violating effect
beyond the standard oscillation picture.

\subsubsection{Case IIb: Same-size source and detector effects with different phases: $|\varepsilon^s_{\alpha}| = |\varepsilon^{d}_{\alpha}|$, $\Delta\phi_{\alpha}\neq 0$\label{sect:IIb}}

An interesting complementary setting is obtained if the magnitude of
the source and detector effects are assumed to be equal so that the
NSI effects can only be distinguished due to the mismatch between
the corresponding CP phases $\phi_{\alpha}^{s}$ and
$\phi_{\alpha}^{d}$. With $|\varepsilon^s_{\alpha}| =
|\varepsilon^{d}_{\alpha}|\equiv |\varepsilon_{\alpha}|$, the
general formula (\ref{eq:Pexpand}) receives a rather symmetric form\footnote{Note  that the coefficient of the last term in Eq.~(\ref{general:sequalsd:raw}) is optically
different from the same in
Eq.~(\ref{eq:Pexpand}) which is due to the utilized goniometric
identity for a difference of two sines and the definition of
$\Delta\phi_{ \alpha}$.}
\begin{eqnarray}
P(\overline{\nu_{e}^{s}}\to \overline{\nu_{e}^{d}}) &\simeq&
P(\overline{\nu_{e}}\to \overline{\nu_{e}})_{\rm SM}\nonumber \\
&-&4\left\{ s_{23}^{2}|\varepsilon_{\mu}|^{2}\cos 2\Delta\phi_{\mu}+
c_{23}^{2}|\varepsilon_{\tau}|^{2}\cos
2\Delta\phi_{\tau}\right.\nonumber \\ &+& \left. 2
c_{23}s_{23}|\varepsilon_{\mu}||\varepsilon_{\tau}|
\cos(\Delta\phi_{\mu}+\Delta\phi_{\tau})\cos
(\Phi_{\mu}-\Phi_{\tau})
\right. \nonumber \\
& +& \left.2s_{13}\left[
s_{23}|\varepsilon_{\mu}|\cos\Delta\phi_{\mu}\cos(\Phi_{\mu}-\delta)+
c_{23}|\varepsilon_{\tau}|\cos\Delta\phi_{\tau}\cos(\Phi_{\tau}-\delta)
\right]
\right\}
\sin^{2}\left(\frac{\Delta m_{32}^{2}L}{4E}\right)\nonumber\\
&-&2\left\{ s_{23}^{2}|\varepsilon_{\mu}|^{2}\sin 2\Delta\phi_{\mu}+
c_{23}^{2}|\varepsilon_{\tau}|^{2}\sin
2\Delta\phi_{\tau}\right.\nonumber\\ &+& \left. 2
c_{23}s_{23}|\varepsilon_{\mu}||\varepsilon_{\tau}|
\sin(\Delta\phi_{\mu}+\Delta\phi_{\tau})\cos
(\Phi_{\mu}-\Phi_{\tau}) \right. \nonumber \\
& +&
\left.2s_{13}\left[
s_{23}|\varepsilon_{\mu}|\sin\Delta\phi_{\mu}\cos(\Phi_{\mu}-\delta)+
c_{23}|\varepsilon_{\tau}|\sin\Delta\phi_{\tau}\cos(\Phi_{\tau}-\delta)
\right] \right\}
\sin\left(\frac{\Delta m_{32}^{2}L}{2E}\right)\nonumber\\
& -&  4 s_{12}c_{12}\left(
c_{23}|\varepsilon_{\mu}|\sin\Delta\phi_{\mu}\cos\Phi_{\mu}-
s_{23}|\varepsilon_{\tau}|\sin\Delta\phi_{\tau}\cos\Phi_{\tau}
\right)\left( \frac{\Delta m_{21}^{2}L}{2E}\right) \,
,\label{general:sequalsd:raw}
\end{eqnarray}
where the notation specified in Eq.~(\ref{notation2}) has been used.
Again, the relevant phase differences in the genuine NSI terms enter in sines.
Furthermore, the formula above can be simplified to a yet more compact form
\begin{eqnarray}
P(\overline{\nu_{e}^{s}}\to \overline{\nu_{e}^{d}}) &\simeq&
P(\overline{\nu_{e}}\to \overline{\nu_{e}})_{\rm SM}\nonumber\\
&-&4\Bigl\{ s_{23}^{2}|\varepsilon_{\mu}|^{2} \sin\left(\frac{\Delta
m_{32}^{2}L}{4E}+2\Delta\phi_{\mu}\right)+
c_{23}^{2}|\varepsilon_{\tau}|^{2}
\sin\left(\frac{\Delta m_{32}^{2}L}{4E}+2\Delta\phi_{\tau}\right)\nonumber\\
&+&
2 c_{23}s_{23}|\varepsilon_{\mu}||\varepsilon_{\tau}|
\sin\left(\frac{\Delta m_{32}^{2}L}{4E}+\Delta\phi_{\mu}+\Delta\phi_{\tau}\right)
\cos (\Phi_{\mu}-\Phi_{\tau})
\nonumber \\
& +& 2s_{13} s_{23}|\varepsilon_{\mu}|\sin\left(\frac{\Delta
m_{32}^{2}L}{4E}+\Delta\phi_{\mu}\right)\cos(\Phi_{\mu}-\delta)\nonumber
\\ & +&2s_{13} c_{23}|\varepsilon_{\tau}|\sin\left(\frac{\Delta
m_{32}^{2}L}{4E}+\Delta\phi_{\tau}\right)\cos(\Phi_{\tau}-\delta)\Bigr\}
\sin\left(\frac{\Delta m_{32}^{2}L}{4E}\right)
\nonumber\\
& -&  4 s_{12}c_{12}\left(
c_{23}|\varepsilon_{\mu}|\sin\Delta\phi_{\mu}\cos\Phi_{\mu}-
s_{23}|\varepsilon_{\tau}|\sin\Delta\phi_{\tau}\cos\Phi_{\tau}
\right) \left(\frac{\Delta m_{21}^{2}L}{2E}\right) \,
,\label{general:sequalsd}
\end{eqnarray}
which does expose the ``kinematic'' role of the phase differences
$\Delta\phi_{\alpha}$ and the ``amplitude modulation'' role of their
averages $\Phi_{\alpha}$.

Let us also remark that for $\Delta\phi_{\alpha}\to 0$ (when the
symmetric setting with $\varepsilon^{s}=\varepsilon^{d\dagger}$
is recovered) the last term vanishes and, as expected, the other
terms conspire to yield a mere shift in the effective  mixing
angle $\tilde\theta_{13}$ identical to that given in formula
(\ref{eq:th13eff}). This provides a nice consistency check of the
results.
A simple numerical analysis of both  Case-IIa and Case-IIb
settings is given in Sect.~\ref{numerics}.

\section{Non-standard antineutrino interactions at Daya Bay \label{numerics}\label{sec:results}}

\subsection{Experimental setting}

The Daya Bay neutrino experiment  \cite{Guo:2007ug} is designed to
perform a precision measurement of $\theta_{13}$ using antineutrinos
produced by the reactors of the Daya Bay Nuclear Power Plant (NPP)
and the Ling Ao NPP. In the detectors, antineutrinos from the
reactors are captured via the inverse beta-decay process, and
the deficit from the expected $1/L^2$ dependence is to be interpreted as a
signature of neutrino oscillations. In particular, near and far
detectors are employed in order to suppress the systematic
uncertainties related to the antineutrino flux from the reactors.

The Daya Bay measurement of $\sin^22\theta_{13}$ is expected to
reach the sensitivity of the order of $0.01$,  an order of
magnitude better than the current CHOOZ limit $\sin^{2}
2\theta_{13}\lesssim 0.17$. Besides a high-quality determination of
the relevant standard neutrino oscillation parameters, Daya Bay can
be rather efficient in improving some of  the current constraints
on physics beyond the SM. 

In order to estimate the NSI effects
possibly observable at Daya Bay, we perform a basic numerical
analysis making use of a simple model of the detected neutrino
spectra.
There are  three pairs of nuclear reactor cores of a total thermal
power of 17.4 GW at the experiment site, namely, Daya Bay (DYB),
Ling Ao (LA) and Ling Ao II (LAII), providing electron antineutrinos
to three detectors, two near ones called Daya Bay (DYB) and
Ling Ao (LA) with 40 tons and a far detector (FAR) with 80 tons of a
Gadolinium-doped liquid scintillator, respectively. A more detailed breakdown of the relevant Daya Bay parameters can be found in  
TABLE~\ref{tab:setting}.
\begin{table}[h]
\begin{tabular}{|c||c|c|c||c|c|}
\hline
DYB layout & \multicolumn{3}{|c||}{geometry} &  \multicolumn{2}{|c|}{expected daily $\overline{\nu}_{e}$ event rates} \\ \hline
Detectors$\backslash$Cores & DYB 2$\times $2.9 GW & LA 2$
\times $2.9 GW & LAII 2$\times $2.9 GW & mods.$\times$DR \cite{Guo:2007ug} & simulated\\ \hline
\hline 
DYB (40~t) &
{363} & 1347 & 1985 & $2\times 930$  & $2\times 890$ \\ \hline LA (40~t) &
{857} & 481 & 1618 & $2\times 760$&  $2\times 790$ \\ \hline FAR (80~t) &
{1307} & 526 & 1613 & $4\times 90$& $4\times 90$ \\ \hline
\end{tabular}\vspace{0.5cm}
\caption{\label{tab:setting}The basic Daya Bay experimental layout \cite{Guo:2007ug} and expected daily antineutrino event rates: distances in meters between detectors (in rows)
and centers of pairs of the neighboring reactor cores (in data columns 1-3) and numbers of anticipated antineutrino events per day at each site (in data columns 4 and 5).  In column 4 we display the data quoted in the DYB proposal (cf. Table 3.8. in \cite{Guo:2007ug}) multiplied by the number of modules, the numbers in column 5 correspond to our numerical analysis described in Section \ref{numerics}. Indeed, the simulated event rates in all cases lay within 4.5\% of the nominal Daya Bay values, thus justifying the relevance of the simplified model of the antineutrino spectrum as well as the expected statistical error levels.}
\end{table}

\subsection{A simple model of the $\overline{\nu_{e}}$ spectra}

For the sake of simplicity, we shall assume that each pair of
neighboring cores constitute a single point source. The average
energy release per one fission $E_{F}$ is anticipated to be around
200 MeV \cite{Guo:2007ug} so the estimated number of fissions per
second in each reactor site $N_{F}$ is
\begin{equation}
N_{F}={2\, P_{T}}/{E_{f}}={ 2\times 2.9~{\rm GW}}/{200~{\rm
MeV}}=1.8\times 10^{20}~s^{-1} \, ,
\end{equation}
where the extra factor 2 counts the number of reactor cores per site
and $P_{T}$ stands for the thermal power of each core. For the
spectrum of the antineutrino flux per fission we shall use the
approximate formula given in Ref.~\cite{Vogel:1989iv} (for $E$ in MeV):
\begin{equation}
\frac{d\Phi }{dE}=\exp \left( 0.87 -0.16 E-0.091 E^{2}\right) {
\rm MeV}^{-1}\,.
\end{equation}

Antineutrinos interact with the free protons in the scintillator via
the inverse beta decay process $\overline{\nu _{e}}+p\rightarrow
n+e^{+}$. The cross-section of this reaction has been calculated in
Ref.~\cite{Vogel:1999zy} to be
\begin{equation}
\sigma \left( E\right) =9.52\times 10^{-48}\left[\left( E-\left(
m_{n}-m_{p}\right) \right) \sqrt{\left( E-\left( m_{n}-m_{p}\right) \right)
^{2}-m_{e}^{2}}\;{\rm MeV}^{-2}\right]{\rm m}^{2}
\end{equation}
with the energy threshold $E_{0}=1.8~$MeV. There are $6.29\times
10^{22}$ free protons in a cm$^{3}$ of the scintillator of density
$\rho =0.86$~g/cm$^{3}$ \cite{nemchenok}. Therefore the number of
targets per one ton of the scintillator is $N_{T}=7.3\times
10^{28}$~ton$^{-1}$.

\begin{figure}[t]
\includegraphics[width=9cm]{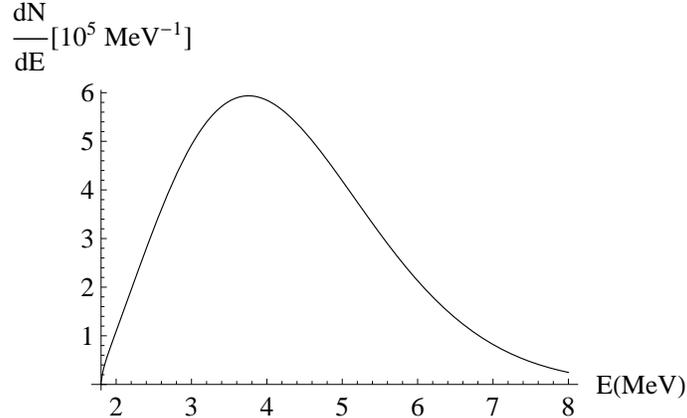}
\caption{\label{fig:DYBspectrum}The expected shape of the detected
antineutrino spectrum in the DYB (near) detector after three years
of running without NSI's.}
\end{figure}

The antineutrino survival probability $P(\overline{\nu_{e}^{s}}\to
\overline{\nu_{e}^{d}})$ is a function of energy, propagation
distance, oscillation parameters and, in general, also the NSI
parameters. The expected total number of antineutrino events in the
detector $D$ (with $D=$DYB,LA,FAR) with mass $M_{D}$ after three
years of running can be estimated as
\begin{eqnarray}
\frac{dN_{D}}{dE} =t\times N_{T}\times M_{D}\times C_{eff} \label{N}
\times N_{F}\times \Phi (E)\times \sigma \left( E\right) \times \sum
_{\substack{{R}={\rm DYB,
LA,LAII}}}\frac{P(\overline{\nu_{e}^{s}}\to
\overline{\nu_{e}^{d}})}{4\pi L_{DR}^{2}} \, ,
\end{eqnarray}
where $t=3\times 365\times 24\times 3600$~s is the duration of a
three-years' run. We sum over three reactor sites and use $L_{DR}$ for
the distance between the detector $D$ and the  reactor site $R$,
c.f. TABLE~\ref{tab:setting}. In addition, we adopt a detection
efficiency coefficient $C_{eff}=0.78$~\cite{Guo:2007ug}. As an
example, we depict in Figure~\ref{fig:DYBspectrum} the expected
spectrum of antineutrinos detected in the DYB detector. It is worth noting that the highest event rate corresponds to $E \simeq
4~{\rm MeV}$.

The quantity of our main interest is the ratio of the
antineutrino energy spectra between the considered far and near
detectors, which can be obtained readily from Eq.~(\ref{N}). It is
expected that, due to a similar design of the far and near
detectors, the systematic uncertainties associated to, e.g., the absolute flux
determination, can be greatly reduced in the ratio of the
energy spectra. However, in order to fully account for all the systematic uncertainties, e.g., the backgrounds, energy miscalibration, detection efficiencies etc., a complex simulation of the Daya Bay experiment is necessary. {This, however, is a formidable task in general, so in what follows we shall consider mainly the statistical uncertainties (Sections \ref{sect:chisquared} and \ref{sect:results}) and only later on (in Section \ref{sect:systematics}) we demonstrate that the changes due to the (leading-order) systematical effects do not inflict any significant changes on these results.}

\subsection{The $\protect\chi^{2}$ analysis\label{sect:chisquared}}

\begin{figure}[t]
\includegraphics[width=8cm]{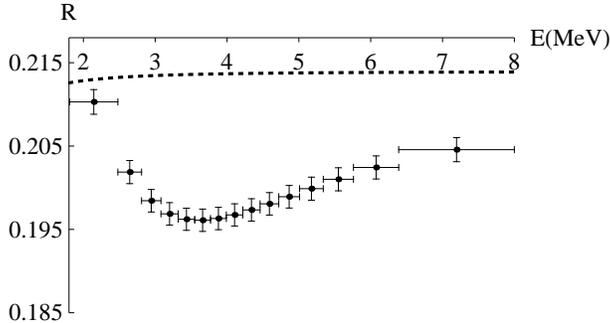}
\caption{\label{fig:pomer}Ratio between the FAR and the DYB
detectors for standard oscillations with statistical errors only. The dotted line corresponds to the case of $\sin^{2} 2\theta_{13}=0$.
The neutrino mixing parameters are the same as those used in
Figure~\ref{fig:fig1}.}
\end{figure}

To assess the observability of NSI's at Daya Bay in practice, we perform a simple numerical
$\chi^{2}$ analysis along the following lines: we choose 15 energy bins from 1.8~MeV to
8~MeV in order to have approximately the same statistics in all bins which are 1-4 times wider than the energy resolution
${15\%}/{\sqrt{E(\text{MeV})}}$ \cite{Guo:2007ug}. In each bin,
we use Eq.~(\ref{N}) to calculate the ratio $R$ of the antineutrino
energy spectra between the far and near detectors (for sake of illustration, from now on we shall focus in particular onto the FAR and the DYB detectors). In the case of
the standard neutrino oscillations, the expected shape of this ratio
between the FAR and the DYB detectors is depicted in
Figure~\ref{fig:pomer}.

Since the
uncertainties of $\theta_{23}$, $\theta_{12}$ and $\Delta
m^{2}_{21}$ are not expected to play any significant role in the
ratio of our interest, we shall fix these parameters to their
central values. This is not the case of $\Delta  m^{2}_{32}$ because
the uncertainty in this parameter (quantified by $\sigma_{\Delta
m^{2}_{32}}$) mimics the effects of the NSI's, namely, it also shifts
the position of the first minimum in $R$.
However, with the increasing precision of the $\Delta m^{2}_{32}$
determination, these effects become less important. Therefore, in
what follows, we shall mainly focus on two specific situation corresponding
to different choices of $\sigma_{\Delta m^{2}_{32}}$: in one case we
take $\sigma_{\Delta m^{2}_{32}}=0.09\times 10^{-3}{\rm eV}^{2}$~\cite{Schwetz:2011qt} as the
current experimental value while in the other ``ideal case'', we
push $\sigma_{\Delta m^{2}_{32}}$ down to $0.025\times 10^{-3}{\rm eV}^{2}$, respectively (which can be viewed as an optimistic expectation for the uncertainty in the atmospheric mass-squared difference in several years from now).

For each specific choice of the relevant NSI parameters there are only two
unknown parameters left in $R$, namely, $s_{13}\equiv \sin\theta_{13}$ and $\Delta m^{2}_{32}$, c.f., Eq.~(\ref{eq:PSM}).
Denoting the $i$-th bin value of $R$ (as a function of $s_{13}$, $\Delta m^{2}_{32}$, $\varepsilon^{s}$ and $\varepsilon^{d}$) by $R_{i}(s_{13},\Delta m^{2}_{32},\varepsilon^{s},\varepsilon^{d})$, we attempt to fit the simulated data by the  NSI null-hypothesis corresponding to the case when $R_{i}$ is calculated for standard oscillations with some effective values of the relevant oscillation parameters, namely, $R^{0}_{i}\left(\tilde{s}_{13},\Delta\tilde{m}^{2}_{32}\right)$. This is done by
minimization of the  $\chi^{2}$ function
\begin{equation}\label{chisq}
\chi ^{2}=\sum_{i=1}^{15}\left[ \frac{R_{i}\left( s_{13},
\Delta m^{2}_{32},\varepsilon^{s},\varepsilon^{d}\right) -R_{i}^{\rm
0}\left( \tilde{s}_{13},\Delta \tilde{m}^{2}_{32}\right)
}{\sigma_{\rm data} }\right] ^{2}+\left(\frac{\Delta
m^{2}_{32}-\Delta
\tilde{m}^{2}_{32}}{\sigma_{\Delta{m}^{2}_{32}}}\right)^{2}
\end{equation}
with respect to $\tilde{s}_{13}$ and $\Delta \tilde{m}^{2}_{32}$, where $\sigma_{\rm data}$ denotes the three-years' run statistical error(s).
The $\tilde{s}_{13}$ parameter has been left free (to be determined  by Daya Bay) whereas $\Delta
\tilde{m}^{2}_{32}$ should obey the existing experimental
constraints. The value of $\chi^{2}$ in the minimum ($\chi^{2}_{\rm min}$) then quantifies
the likelihood that the Daya Bay data
could be fitted by the standard oscillation formula. For two fitted parameters, $\chi^{2}_{\rm min}=2.3$ and $4.61$ corresponds to $68\%$ and $90\%$ C.L., respectively. 

\subsection{Results\label{sect:results}}

For the sake of simplicity, in what follows we shall consider only the ``flavourless''
versions of the oscillation probability formulas relevant to the
three cases of our interest, namely Eqs.~(\ref{eq:PcaseI}),~(\ref{general:sonly:raw}) and~(\ref{general:sequalsd}). This amounts to setting
$\phi_{\mu}=\phi_{\tau}\equiv \phi$ and
$|\varepsilon_{\mu}|=|\varepsilon_{\tau}|\equiv |\varepsilon|$ everywhere.
Let us recall that, besides the standard oscillation parameters, in
Case I and Case IIa the relevant input NSI parameters are, namely, the
(universal) magnitude of the NSI effects $|\varepsilon|$ and the
corresponding CP phase $\phi$ (more precisely, the phase difference
$\phi'=\phi-\delta$ where $\delta$ denotes the leptonic Dirac CP
phase) while in Case IIb the NSI parameters entering the survival
probability are $|\varepsilon|$, $\Phi$ (again, it is rather
$\Phi'=\Phi-\delta$) and $\Delta\phi$.

Let us also reiterate that only the statistical errors have been taken
into account in the current analysis. A complete study including
also the systematic uncertainties would require a complex simulation
of the Daya Bay experiment. This, however, is beyond the scope of
the present study.

\subsubsection{Case I}

\begin{figure}[t]
\includegraphics[width=8cm]{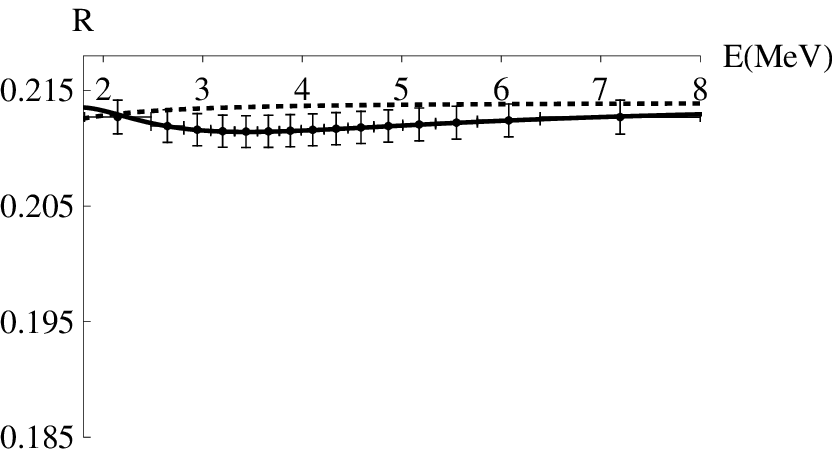}
\includegraphics[width=8cm]{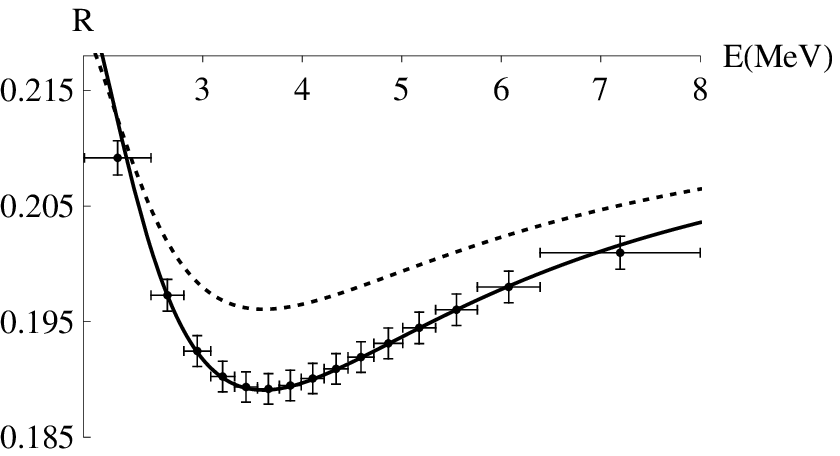}
\caption{\label{plots12} Sample fits of the ratio between the
detected antineutrino spectra in the FAR and DYB near detectors in
the case that the production and detection process (including the relevant non-standard interactions) are just inverse of each other (Case I, Sect.~\ref{sect:CaseI}), i.e., $\varepsilon^{s}=\varepsilon^{d\dagger}$. In the left panel,
$\sin ^{2}2\protect\theta _{13}=0 $, $\left\vert \protect\varepsilon
\right\vert =0.04$ and $\phi'=0$ have been used; in the right panel,
$\sin ^{2}2\protect\theta _{13}=0.1$, $\left\vert
\protect\varepsilon \right\vert =0.02$ and $\phi'=0$ instead. The
dotted lines correspond to the standard oscillations without NSI's,
while the solid lines are the fits based on the standard oscillation
survival probability (\ref{eq:PSM}) used in Eq.~(\ref{N}) with the
effective mixing angles given by $\sin ^{2}2\tilde{\protect\theta
}_{13}=0.013$ (left panel) and $\sin ^{2}2\tilde{\protect\theta
}_{13}=0.138$ (right panel).}
\end{figure}
As we argued in Sect.~\ref{sect:CaseI}, in the symmetric setting
with $\varepsilon ^{s}=\varepsilon ^{d\dagger }$, the NSI effects
cannot be distinguished from the pure standard oscillations.
Even if the underlying mixing angle $\theta _{13}$ is zero, one can
still fit the data with a standard oscillation curve corresponding
to a nonzero value of the effective mixing angle $\tilde\theta
_{13}$ given by formula (\ref{eq:th13eff}). A pair of representative
plots depicting the expected ratio between the FAR and the DYB
antineutrino spectra in this situation are given in
Figure~\ref{plots12}.
One can see that the data are well fitted by the
standard oscillation formula with just the effective mixing angles different from their ``true'' values.

\subsubsection{Case IIa\label{num:caseIIa}}

In the ``flavourless'' setting, the relevant Case-IIa formula
(\ref{general:sonly:raw}) simplifies into
\begin{eqnarray}
P(\overline{\nu_{e}^{s}}\to \overline{\nu_{e}^{d}}) &\simeq&  P(\overline{\nu_{e}}\to \overline{\nu_{e}})_{\rm SM}
\nonumber-4s_{13}
(s_{23}+c_{23})|\varepsilon|\cos \phi'
\sin^{2}\left(\frac{\Delta m_{32}^{2}L}{4E}\right)\nonumber\\
& +& 2 s_{13}(s_{23}+c_{23})|\varepsilon|\sin\phi'
\sin\left(\frac{\Delta m_{32}^{2}L}{2E}\right)\nonumber\\
& +&  2 s_{12}c_{12}
(c_{23}-s_{23})|\varepsilon|\sin\phi\left(\frac{\Delta m_{21}^{2}L}{2E}\right)\,,
\label{flavourblindcaseIIa}
\end{eqnarray}
where $\phi'\equiv \phi-\delta$. Note that in most cases the last
term can be neglected due to the experimental proximity of
$\theta_{23}$ to $\frac{\pi}{4}$.

\begin{figure}[t]
\includegraphics[width=8cm]{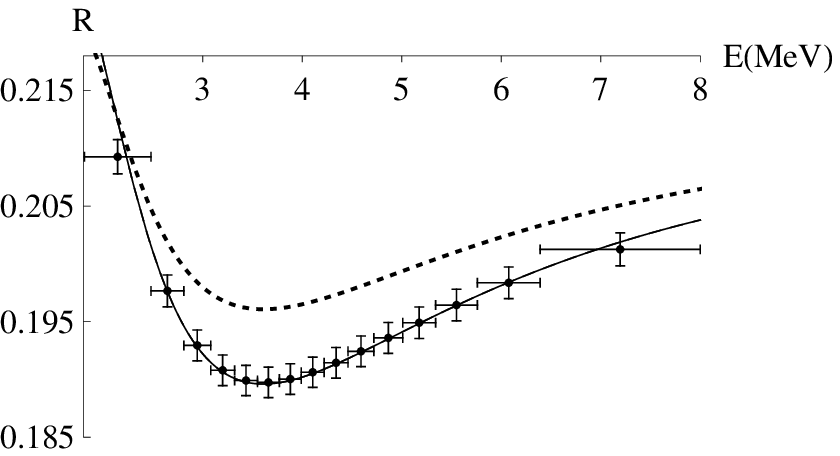}
\includegraphics[width=8cm]{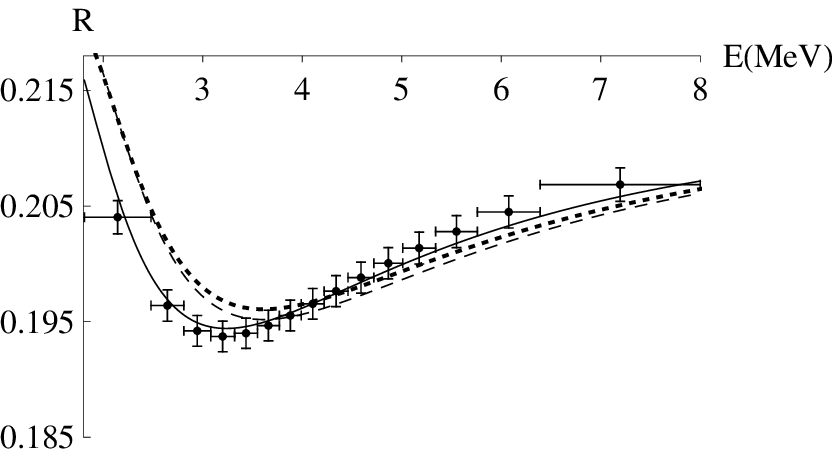}
\caption{\label{sourceonlyvariousphiprimes} Sample fits of the ratio
between the detected antineutrino spectra in the FAR and DYB near
detectors in the asymmetric setting where the NSI's are assumed
to affect only the production process (Case IIa,
Sect.~\ref{sect:caseIIa}). We adopt $\sin ^{2}2 \theta _{13}=0.1$
and $|\varepsilon| =0.04$ in both panels. Furthermore, in the left
panel, $\phi'=0$ is assumed, while in the right panel 
we take $\phi'=\frac{\pi}{2}$ (maximal CP phase difference). The
dotted lines correspond to the standard oscillations without NSI's,
while the dashed lines show the fitted curves with the effective
mixing angle $\sin^{2}2 \tilde\theta_{13}=0.135$ (left panel) and
$\sin^{2}2 \tilde\theta_{13}=0.105$ (right panel). In addition, the
solid lines stand for the fitted curves with two parameters
$\tilde\theta_{13}$ and $\Delta \tilde m^2_{32}$. In the left panel,
the solid line coincides with the dashed line, whereas in the right panel it does not and the best fit 
corresponding to the values $\sin^{2}2
\tilde\theta_{13}=0.109$ and $\Delta \tilde m^2_{32} = 2.20\times
10^{-3}~{\rm eV}^2$ requires a significant shift in $\Delta \tilde{m}^{2}_{32}$ with respect to its central value.}
\end{figure}

We also stress that for $\phi'\to 0$ or $\pi$ the leading NSI
contribution corresponding to the sine-squared term above essentially mimics the
effects of standard oscillations with a shifted mixing angle
because, besides the last negligible term, there is no net NSI
induced CP-violating effect. One can see this on the left panel in
Figure \ref{sourceonlyvariousphiprimes} where, indeed,  the data can
be fitted by the standard oscillation formula with just a shifted
effective mixing angle $\tilde\theta_{13}$.
However, the change is still proportional to $s_{13}$ and thus no
shift is induced if the underlying $\theta_{13}$ happens to be zero\footnote{As we shall see, this is different from the Case IIb setting studied
in Sect.~\ref{num:caseIIb} where a nonzero value of the effective
mixing angle can be generated even for $\theta_{13}=0$.}.

However, for non-trivial $\phi'$, the NSI effects can no longer be
subsumed into a pure shift in $\theta_{13}$ and the standard
oscillation formula no longer fits the data even if one admits for a
certain variation in $\Delta m^{2}_{32}$, see the right panel in
Figure~\ref{sourceonlyvariousphiprimes}. Thus, in this case, one can in principle attempt to constrain the $|\varepsilon|$ and
$\phi'$ parameters, at least in some parts of their parameter space.

\begin{figure}[t]
\includegraphics[width=8cm]{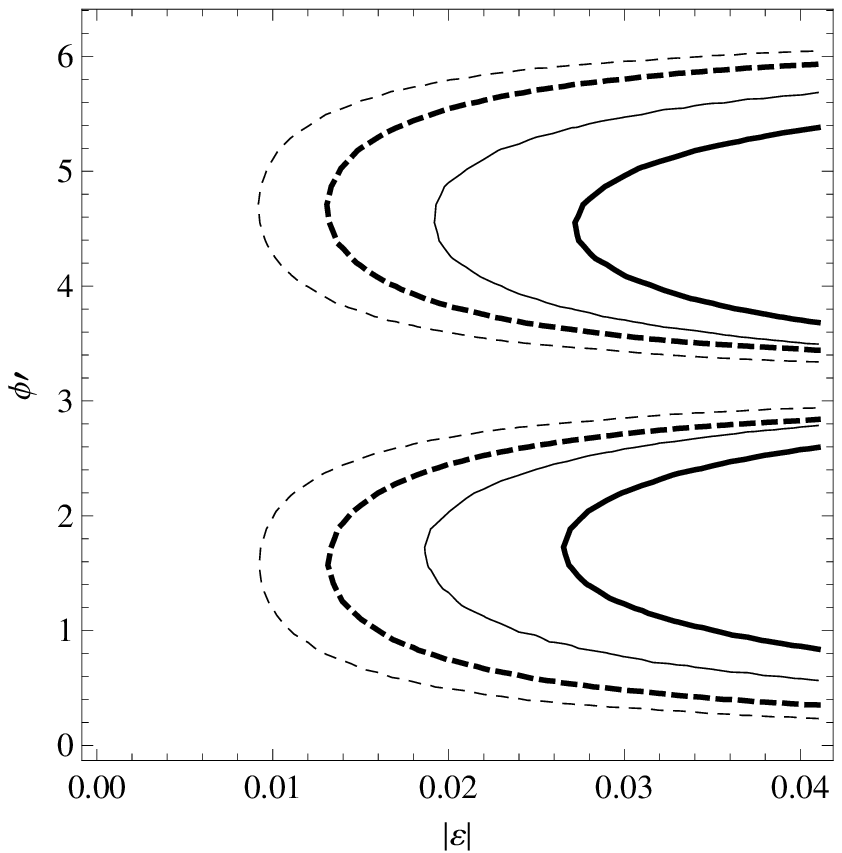}
\caption{\label{sourceonlycontour} Regions of parameters (on the right hand side
of the curves),  where the Daya Bay experiment can disfavour the standard oscillation hypothesis at $68\%$ (thin curves) and at $90\%$ (thick curves) C.L.
for $\sin ^{2}2\protect\theta _{13}=0.1$. The solid curves stand for a standard two-parameter fit
with variable $\theta_{13}$ and  $\Delta m^2_{32}$ for $\sigma_{\Delta m^{2}_{32}}=0.09\times 10^{-3}{\rm eV}^{2}$ while the 
dashed curves correspond to the fit with $\sigma_{\Delta m^{2}_{32}}$ pushed down to $0.025\times 10^{-3}{\rm eV}^{2}$.}
\end{figure}

In Figure~\ref{sourceonlycontour} we present the relevant exclusion regions for these parameters.
Therein, one can observe an interesting $\pi$-periodicity in
$\phi'$, which can be understood from the shape of the second
correction in formula (\ref{flavourblindcaseIIa}). Remarkably enough, even with variable $\Delta m^{2}_{32}$ and $|\varepsilon|$ as low as 0.02, in some cases the NSI effects can be distinguished from the standard oscillation at $90\%$ C.L. 

\subsubsection{Case IIb\label{num:caseIIb}}

In the more general case when both the source and detector effects
are present there is an extra set of parameters at play associated
to $\varepsilon^{d}_{\alpha}$, i.e., the relevant magnitudes
$|\varepsilon^{d}_{\alpha}|$ and also the extra detector NSI phases
$\phi^{d}_{\alpha}$ which combine with the source ones into the
phase averages $\Phi_{\alpha}$ and the phase differences
$\Delta\phi_{\alpha}$, c.f. Eq.~(\ref{notation2}). As before, we
will assume a ``flavourless'' form of NSI's and deliberately put
$|\varepsilon^{d}|=|\varepsilon^{s}|\equiv |\varepsilon|$ in order
to simplify the numerical analysis. Then formula
(\ref{general:sequalsd:raw}) reduces to
\begin{eqnarray}
P(\overline{\nu_{e}^{s}}\to \overline{\nu_{e}^{d}}) &\simeq&  P(\overline{\nu_{e}}\to
\overline{\nu_{e}})_{\rm SM}\nonumber \\ &-& 4
(s_{23}+c_{23})^{2}|\varepsilon|^{2} \sin\left(\frac{\Delta
m_{32}^{2}L}{4E}\right) \sin\left(\frac{\Delta
m_{32}^{2}L}{4E}+2\Delta\phi\right)
\nonumber \\
&-& 8s_{13}
(s_{23}+c_{23})|\varepsilon|\cos \Phi'
\sin\left(\frac{\Delta m_{32}^{2}L}{4E}\right)
\sin\left(\frac{\Delta m_{32}^{2}L}{4E}+\Delta\phi\right)\nonumber\\
& -&  4 s_{12}c_{12} (c_{23}-s_{23})|\varepsilon|\sin\Delta\phi
\cos\Phi \left(\frac{\Delta m_{21}^{2}L}{2E}\right) \, ,
\label{flavourblindcaseIIb}
\end{eqnarray}
where, again, $\Phi'\equiv \Phi-\delta$. As before, the last term is
negligible for $\theta_{23}$ close to $\pi/4$.

\begin{figure}[t]
\includegraphics[width=8cm]{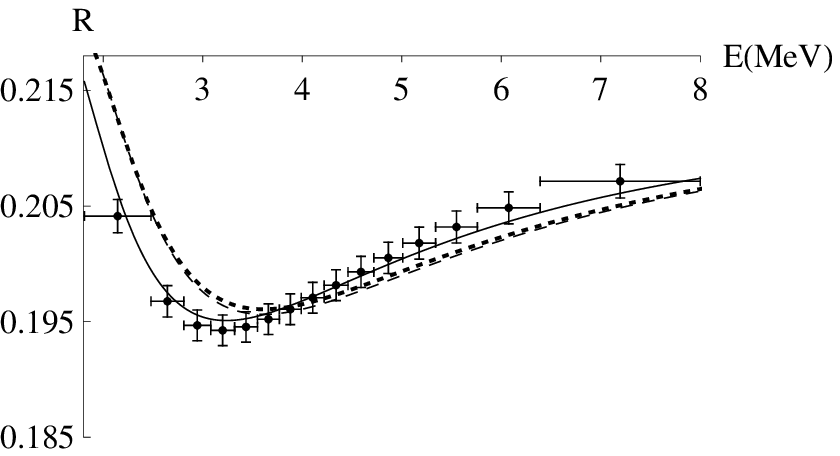}
\includegraphics[width=8cm]{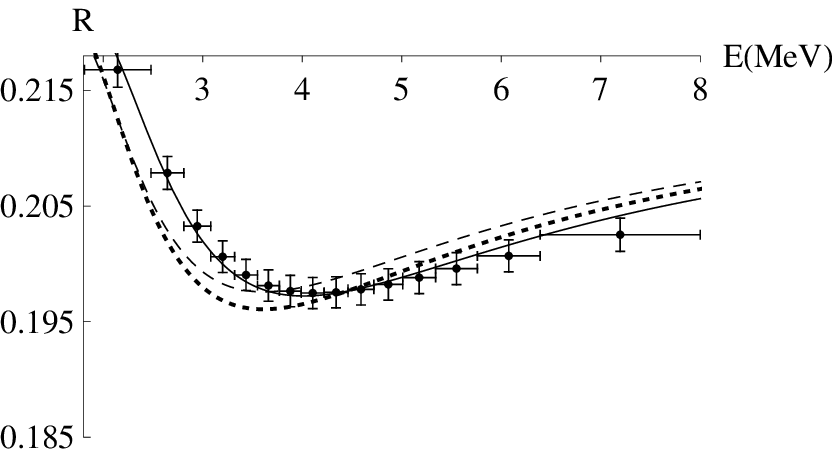}
\caption{\label{plots34} Sample fits of the ratio between the
detected antineutrino spectra in the FAR and DYB near detectors in
Case IIb. We adopt $\sin ^{2}2 \theta _{13}=0.1$ and $|\varepsilon|
=0.02$ as well as $\Delta\phi=\frac{\pi}{2}$ in both panels.
Furthermore, in the left panel, $\Phi'=\frac{\pi}{2}$ is assumed,
while in the right panel we put $\Phi'=\frac{3}{2}\pi$. 
The dotted lines correspond to the standard
oscillations without NSI's, while the dashed lines show the fitted
curves with the effective mixing angle $\sin^{2}2
\tilde\theta_{13}=0.103$ (left panel) and $\sin^{2}2
\tilde\theta_{13}=0.092$ (right panel). In addition, the solid lines
stand for the fitted curves with two parameters, i.e., $\sin^{2}2
\tilde\theta_{13}=0.105$ and $\Delta \tilde m^2_{32} = 2.20\times
10^{-3}~{\rm eV}^2$ in the left panel, and $\sin^{2}2
\tilde\theta_{13}=0.094$ and $\Delta \tilde m^2_{32} = 2.72\times
10^{-3}~{\rm eV}^2$ in the right panel.}
\end{figure}

For the sake of illustration, in Figure~\ref{plots34} we show two specific examples of the $R$-fits obtained in Case IIb. There, the data are fitted by the standard oscillations, first with variable $\theta_{13}$ and $\Delta m^2_{32}$ (solid lines) and then also with only $\theta_{13}$ as a free parameter (dashed lines).

\begin{figure}[t]
\includegraphics[width=8cm]{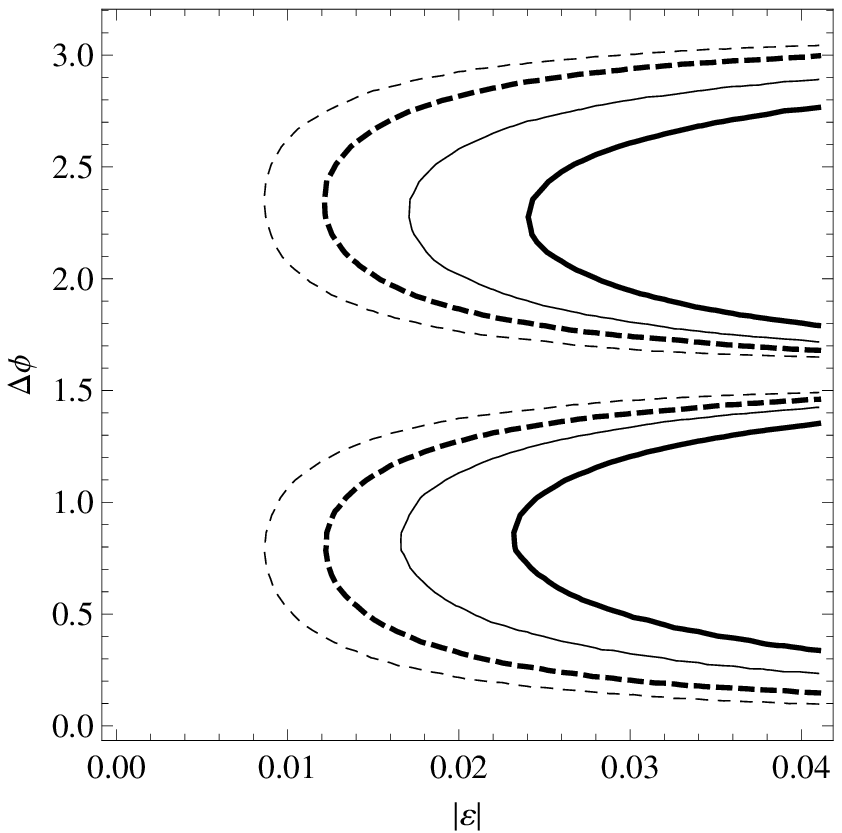}
\includegraphics[width=8cm]{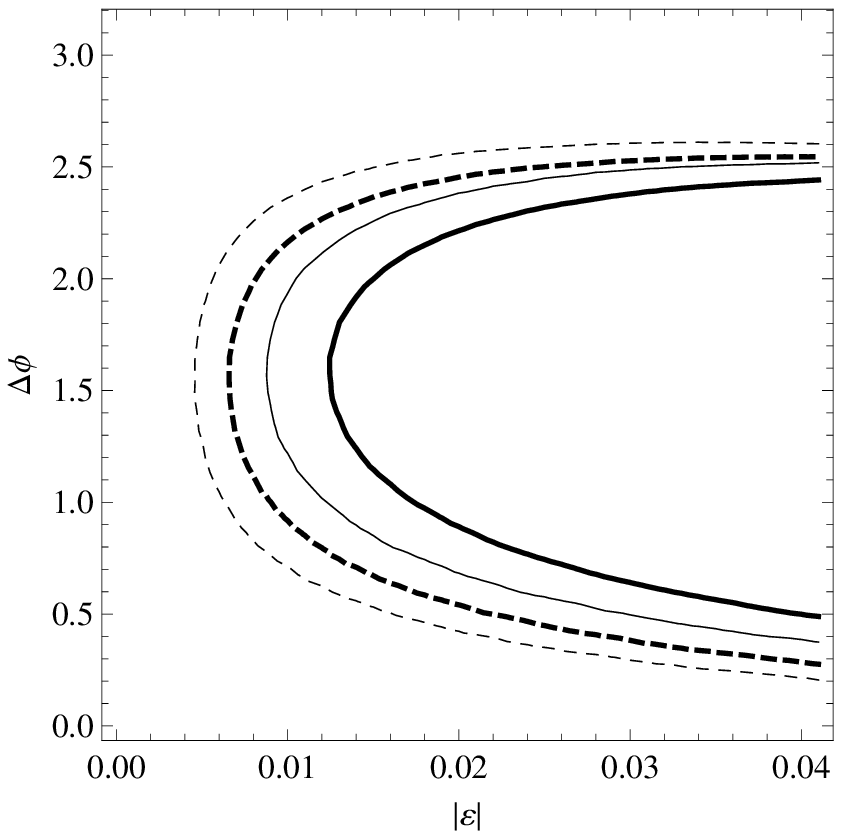}
\caption{\label{epspsi} Regions of parameters (on the right hand side
of the curves),  where the Daya Bay experiment can disfavour the standard oscillation hypothesis at $68\%$ (thin curves) and at $90\%$ (thick curves) C.L.
for $\sin ^{2}2\theta_{13}=0.1$. Here we assume
$\Phi'=0$ (left panel) and $\Phi'=\frac{\pi}{2}$ (right panel). As before, the solid curves stand for a standard two-parameter fit
with variable $\theta_{13}$ and $\Delta m^2_{32}$ for $\sigma_{\Delta m^{2}_{32}}=0.09\times 10^{-3}{\rm eV}^{2}$ while the 
dashed curves correspond to the fit with $\sigma_{\Delta m^{2}_{32}}$ pushed down to $0.025\times 10^{-3}{\rm eV}^{2}$.}
\end{figure}

In Figure~\ref{epspsi}, the exclusion plots for the $\Delta\phi$ and $|\varepsilon|$ parameters are given for $\sin^{2}2\theta_{13}$ and two specific choices of $\Phi'$.
The sensitivity in $|\varepsilon|$ is similar to that observed in 
Figure~\ref{plots34} for Case IIa. Notice, however, that the two leading corrections in
Eq.~(\ref{general:sequalsd:raw}) have a very different
$\Delta\phi$-periodicity. The former is $\pi$-periodic in
$\Delta\phi$ while the latter is $\frac{\pi}{2}$-periodic in
$\Delta\phi$. The reason is easily seen from the analytic shape of the relevant survival probability (\ref{flavourblindcaseIIb}). Indeed, for $\Phi'=0$, the second correction in
formula (\ref{flavourblindcaseIIb}) dominates over the first one
while it is the other way round for $\Phi'=\frac{\pi}{2}$.

\begin{figure}[t]
\includegraphics[width=7.5cm]{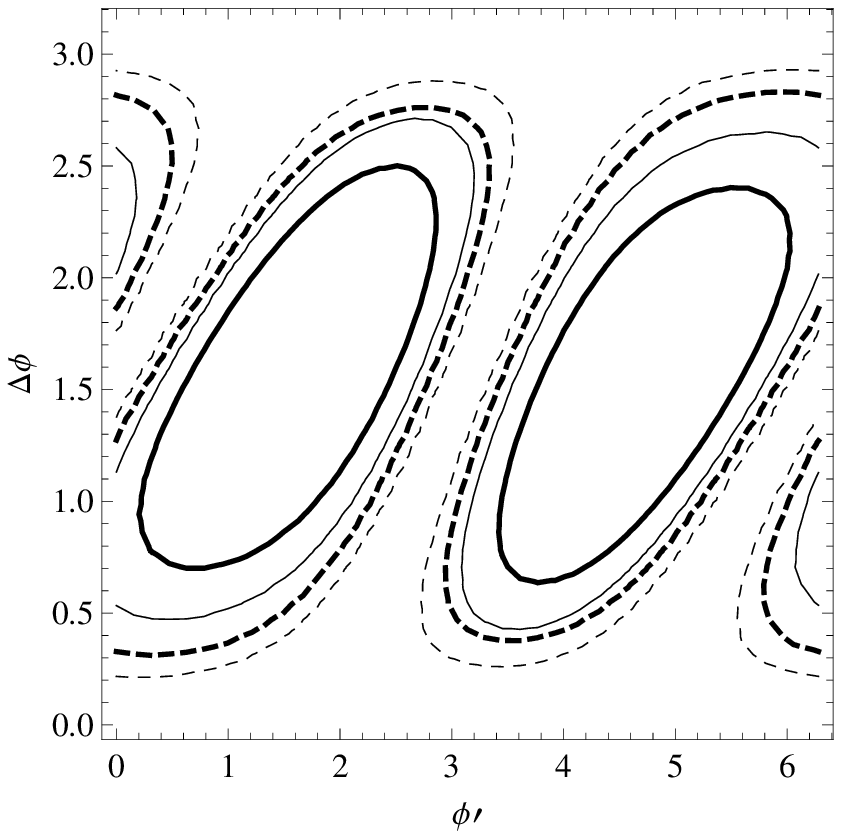}
\includegraphics[width=7.5cm]{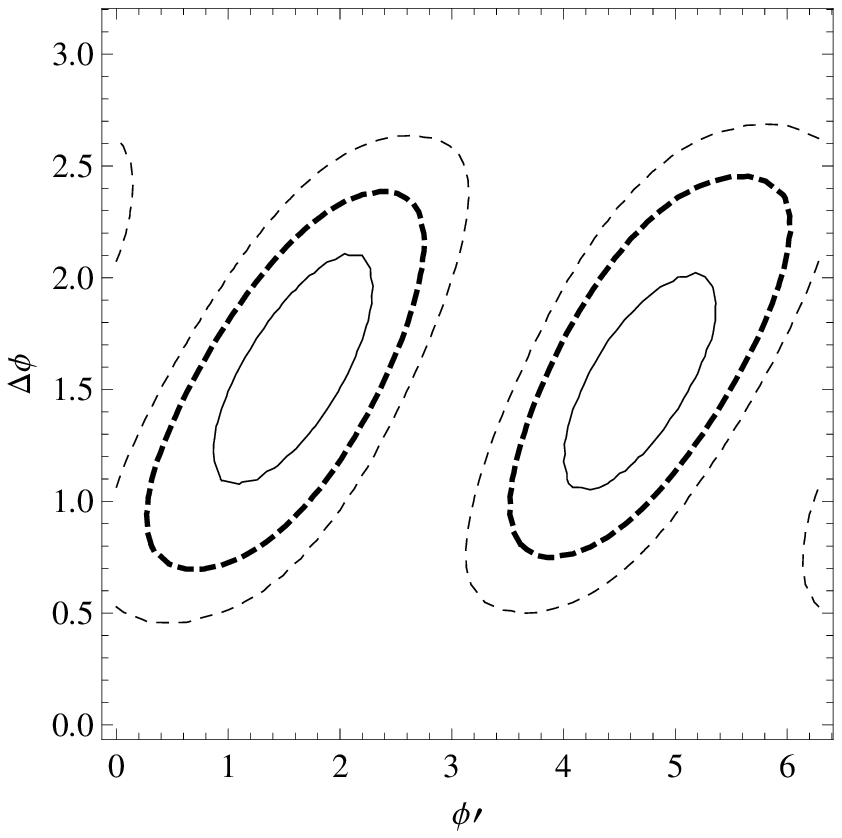}\\
\vspace{0.5cm}
\includegraphics[width=7.5cm]{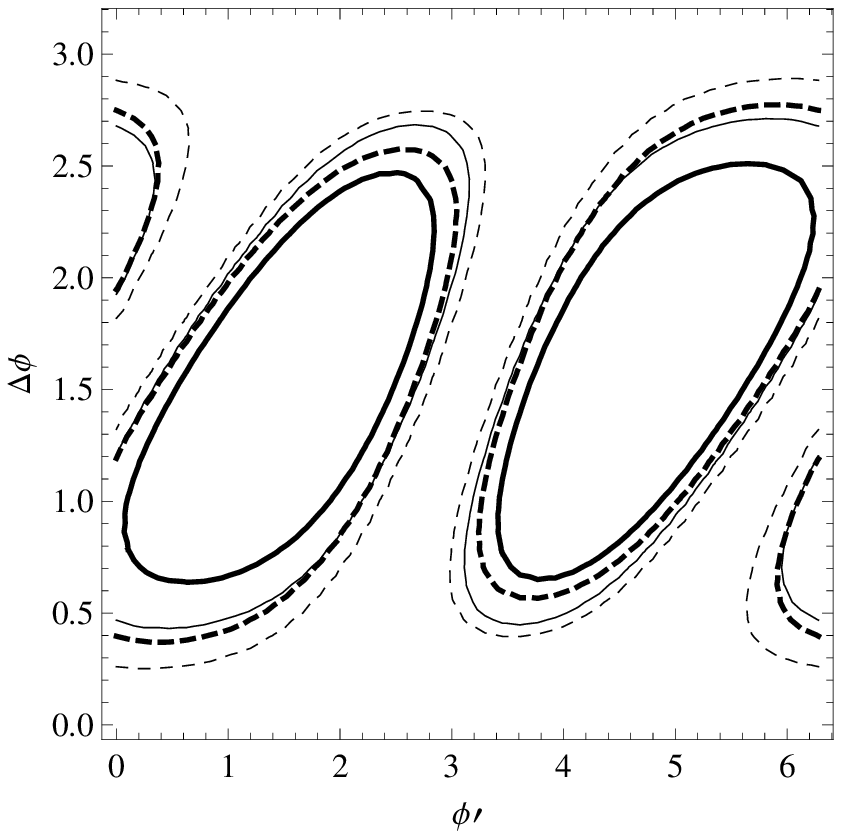}
\includegraphics[width=7.5cm]{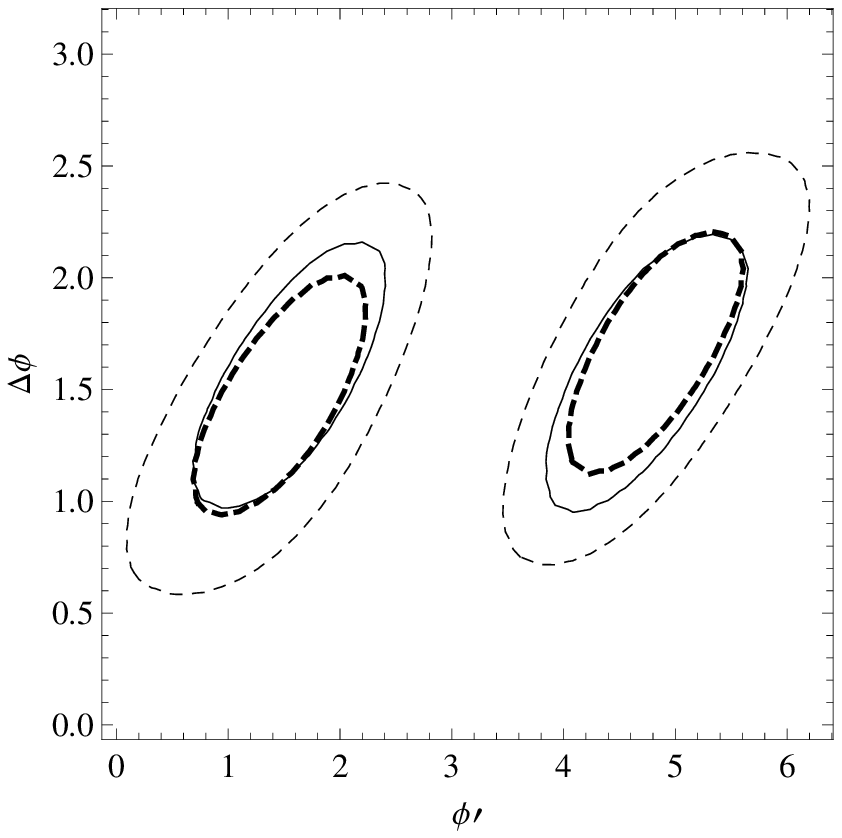}
\caption{\label{contours1}Regions of parameters (interior),  where the Daya Bay experiment can disfavour the standard oscillation hypothesis at $68\%$ (thin curves) and at $90\%$ (thick curves) C.L. for $\left\vert
\varepsilon \right\vert =0.02$ (left column) and $\left\vert
\varepsilon \right\vert =0.01$ (right column) and $\sin
^{2}2\theta_{13}=0.1$ (upper row) and $\sin ^{2}2\theta _{13}=0.05$
(lower row). Again, the solid curves stand for a standard two-parameter fit
with variable $\theta_{13}$ and $\Delta m^2_{32}$ for $\sigma_{\Delta m^{2}_{32}}=0.09\times 10^{-3}{\rm eV}^{2}$ while the 
dashed curves correspond to the fit with $\sigma_{\Delta m^{2}_{32}}$ pushed down to $0.025\times 10^{-3}{\rm eV}^{2}$.}
\end{figure}

Regions in the $\Phi' - \Delta\phi $ plane where Daya Bay experiment
could distinguish non-standard effects from standard oscillations
(at $68\%$ and $90\%$ C.L.) are shown in Figure~\ref{contours1} for
different values of $\left\vert \varepsilon \right\vert $ and
$\sin^{2}2\theta _{13}$.
If  the value of the underlying $\theta _{13}$ is close to the CHOOZ limit
($\sin^22\theta_{13}<0.17$) and $\left\vert \varepsilon \right\vert
=0.02$ then the region is relatively large, see the upper-left panel
in Figure~\ref{contours1}. With decreasing $\left\vert \varepsilon
\right\vert $ (from left to right) or $\theta _{13}$ (from up to
down), the observability domains become naturally smaller.

\begin{figure}[t]
\includegraphics[width=8.5cm,height=8cm]{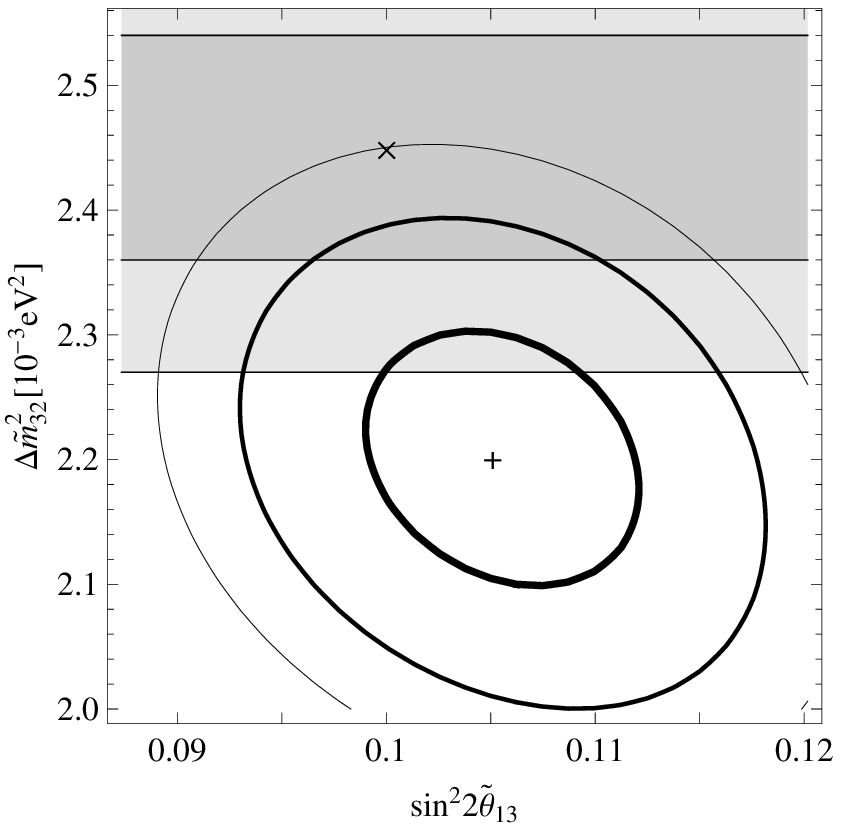}
\caption{\label{thetadelta}The effects of the non-standard interactions in the determination of the standard oscillation parameters $\theta_{13}$ and $\Delta m^{2}_{32}$ at Daya Bay after 3 years of running. The upper cross denotes the assumed ``true'' values of the standard oscillation parameters $\sin^{2} 2\theta_{13}=0.1$, $\Delta m^{2}_{32}=2.45~\times~10^{-3}$eV$^{2}$.
Turning on the NSI parameters (fixing, for instance, $|\varepsilon|=0.02$,  $\Phi=\pi/2$, $\Delta\phi=\pi/2$ in case IIb, c.f., Sect.~\ref{sect:IIb}), the best standard oscillation fit is shifted to $\sin^{2} 2\tilde \theta_{13}=0.105$ and $\Delta \tilde m^{2}_{32}=2.20~\times~10^{-3}$eV$^{2}$ (the lower cross) and the corresponding $\chi^{2}_{\rm  min}=12.6$ indicates a significant incompatibility between the Daya Bay data and the standard oscillation hypothesis. We display three solid curves depicting the $\chi^{2}$ levels around the best-fit point; from thick to thin, $\chi^{2}=20, 40$ and~$60$, respectively. The shaded bands depict the pull (due to the second term in Eq.~(\ref{chisq})) inflicted by  $\Delta \tilde{m}^{2}_{32}$ departing from the ``true value''; the dark/light and light/white boundaries enclose the  $\Delta \tilde{m}^{2}_{32}=(2.45\pm 0.09) \times 10^{-3}{\rm eV}^{2}$ and $\Delta \tilde{m}^{2}_{32}=(2.45\pm 0.18) \times 10^{-3}{\rm eV}^{2}$ regions (about $1\sigma$ and $2\sigma$), respectively.   
}\end{figure}

The possible NSI effects in an independent Daya Bay determination of the standard oscillation parameters  $\theta_{13}$ and $\Delta m^{2}_{32}$ are illustrated in Figure~\ref{thetadelta}. One can see that, at least in some cases, the corresponding global best fit point can differ significantly from the ``true'' values of these parameters, potentially leading to a tension between Daya Bay and other experiments.   

Yet another comment is in order here. As we have seen in Case IIa
(c.f., Sect.~\ref{num:caseIIa}), with source effects only there is
no way to end up with a significant effective $\theta_{13}$ if the
underlying $\theta_{13}$ was zero, while here one still gets
$\tilde\theta_{13}\neq 0$ even for $\theta_{13}=0$ due to the first
term in Eq.~(\ref{flavourblindcaseIIb}). Such a qualitative
difference in the behavior of these two settings can be
heuristically understood as follows: In the former case, there are
effectively only two small parameters (with their corresponding CP
phases) at play, namely $|\varepsilon^{d}|$ and $s_{13}$ while there
are three such quantities in the latter case, in particular
$|\varepsilon^{s}|$, $|\varepsilon^{d}|$ and $s_{13}$. In Case IIa,
there is thus only a single relevant phase difference governing the
CP-even effects [due to the first correction in
Eq.~(\ref{flavourblindcaseIIa})] which, however, becomes ill defined
in the $s_{13}\to 0$ limit, and thus its effect can be
``rotated away''. Remarkably, this is not so in Case IIb since there
is an observable phase difference $\Delta\phi$ left even in the
$s_{13}\to 0$ limit and the corresponding contribution to the
effective $\tilde\theta_{13}$ due to the first term in
Eq.~(\ref{flavourblindcaseIIb}) cannot be transformed out.

{
\subsection{Effects of leading-order systematics\label{sect:systematics}}
Finally, let us argue that the leading systematical effects do not change the results obtained in the previous sections in any significant way. 

The main sources of systematical uncertainties in the Daya Bay setting are related to the reactors (power, spent fuel, location), detectors (energy miscallibration, target mass, detector efficiency) and, of course, backgrounds (accidental signals, $^{8}$He $^{9}$Li, fast neutrons)~\cite{Guo:2007ug}. Taking the full advantage of the ``near+far'' detector setting
one can approximate the leading-order systematic uncertainties (namely, the neutrino flux uncertainty and the uncertainty in the detector masses) as a relative change of the measured far-to-near ratio of the detected antineutrino
energy spectra $R$, cf. Section \ref{sect:chisquared}. For the sake of simplicity we shall assume that this change (to be denoted by $K$), as well as its uncertainty $\sigma_{K}$, are energy-independent at the leading order. Following the detailed discussion given in \cite{Guo:2007ug} we shall adopt a conservative value of $\sigma_{K}= 0.6\%$  for the calculation.

The argument above makes it possible to implement the leading-order systematics by simply extending the original formula (\ref{chisq}) into
\begin{eqnarray}
{ \chi }^{2} &=&\sum_{i=1}^{15}\left[ \frac{R_{i}\left(
s_{13},\Delta m_{32}^{2},\varepsilon ^{s},\varepsilon ^{d}\right) -K\times
R_{i}^{SM}\left( \widetilde{s}_{13},\Delta \widetilde{m}_{32}^{2}\right) }{%
\sigma _{data}}\right] ^{2} \label{ChiSquare} \\ 
&+&\left( \frac{\Delta m_{32}^{2}-\Delta \widetilde{m}_{32}^{2}}{%
\sigma _{\Delta m_{32}^{2}}}\right) ^{2}+\left( \frac{K-1}{\sigma _{K}}
\right) ^{2}  \nonumber
\end{eqnarray}
and marginalizing over $s_{13}$, $\Delta \widetilde{m}_{32}^{2}$ and $K$. With three degrees of freedom at play the 68\% C.L. and 90\% C.L.  
values now correspond to  $\chi ^{2}=3.53$  and $\chi ^{2}=6.25$, 
respectively.

Let us illustrate the smallness of the changes inflicted by the variation of $K$ on, e.g., the situation of  
Case IIb studied in detail in Section \ref{num:caseIIb}.  In Figure \ref{fig:systematics} we demonstrate the shift in the solid contours displayed previously in Figure \ref{epspsi} (where these were obtained for $\sigma_{\Delta m^{2}_{32}}=0.09\times 10^{-3}{\rm eV}^{2}$ with only statistical uncertainties taken into account) due to the systematic effects. The dashed lines in Figure \ref{fig:systematics} demonstrate the slightly reduced sensitivity of the Daya Bay if the leading systematics is taken into account. 
To conclude, the systematics does not hinder the Daya Bay's sensitivity of to the new physics effects and the discovery reach remains safely statistics-dominated. 

\begin{figure}[h]
\centering
\includegraphics[width=7.5cm]{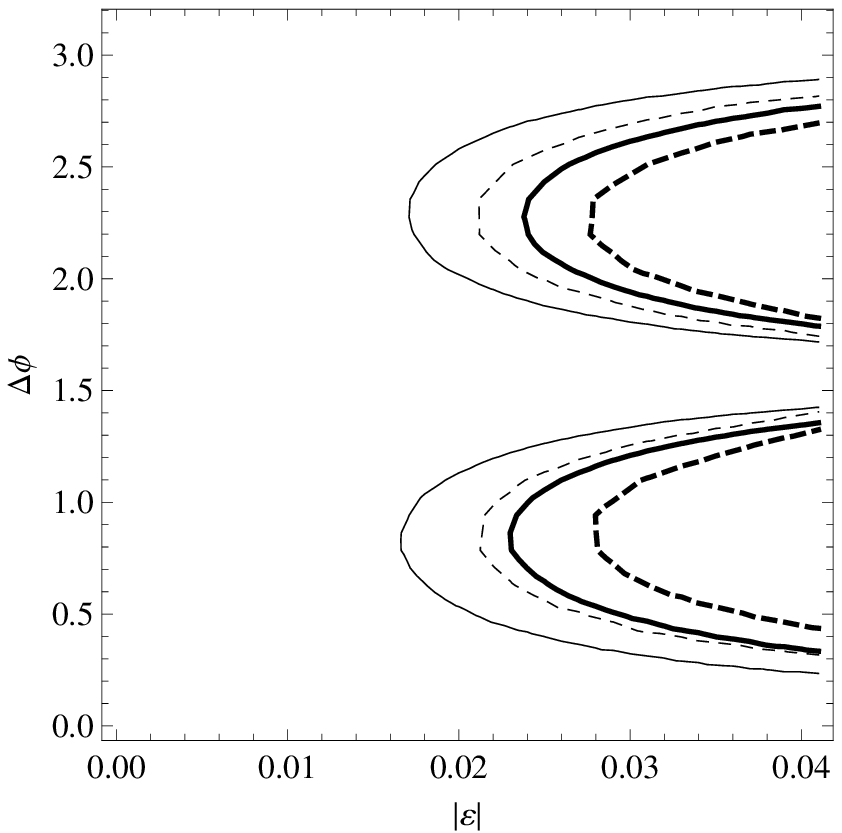}
\includegraphics[width=7.5cm]{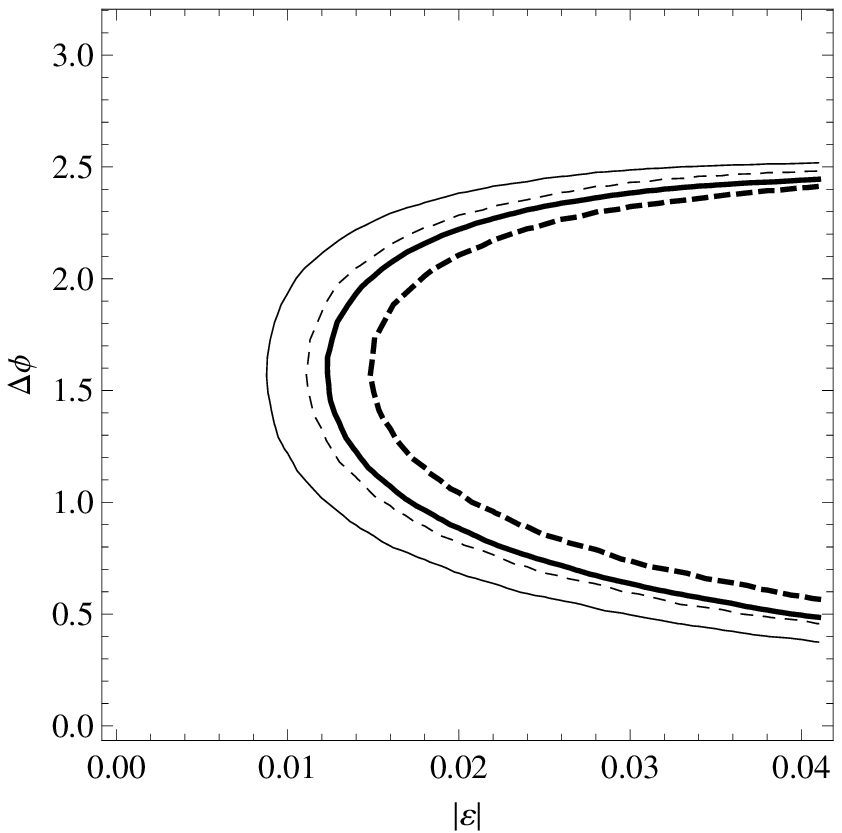}
\caption{The Daya Bay new-physics sensitivity with both statistics and systematics taken into account (in dashed lines) as compared to the results obtained previously with solely statistical uncertainties (in solid lines), cf. Figure \ref{epspsi}. As usual, the thick and thin lines correspond to the $90\%$ C.L. and $68\%$ C.L. contours, respectively. All the NSI and oscillation parameters are fixed to the values employed in  Figure \ref{epspsi}, i.e., $\sin ^{2}2\theta_{13}=0.1$, $\Phi'=0$ (left panel) and $\Phi'=\frac{\pi}{2}$ (right panel), with  $\sigma_{\Delta m^{2}_{32}}=0.09 \times 10^{-3}$~eV$^2$.}
\label{fig:systematics}
\end{figure}
}
\section{Summary and outlook}
\label{sect:summary}

In this work, we have performed a detailed analysis of the non-standard
antineutrino interaction effects in the Daya Bay short-baseline
reactor antineutrino experiment.

The NSI's in reactor antineutrino experiments can exhibit themselves
in various ways depending on the character of the underlying
physics. If, for instance, the non-standard interactions in the production and detection processes
happen to be exactly the same, i.e., $\varepsilon^{s}=\varepsilon^{d\dagger}$, the net effect
consists in a shift in the depth of the oscillation dip in the
measured ratio of the far and near detector antineutrino spectra
corresponding to the extracted value of the mixing angle
$\theta_{13}$. Thus, in this case, the NSI effects can not be
distinguished from the standard oscillations~\cite{Ohlsson:2008gx}.

If, however, this assumption is relaxed, owing to, e.g.,
non-standard multi-body interactions in the source, the measured
antineutrino spectra are distorted in a specific way and become
incompatible with the standard oscillation interpretation -- besides
the change of the depth of the first dip, also its energy position
is shifted. This can be only partially accounted for by the standard
oscillation formula if the extracted values of the mixing angle
$\theta_{13}$ and, in particular, the corresponding mass-squared
difference $\Delta m^2_{32}$, are both allowed to differ
significantly from their genuine values. However, in practice, the
effect can not be entirely subsumed into a shift in the
$\theta_{13}-\Delta m^{2}_{32}$ plane due to the strict constraints
on these parameters from other measurements.

In Sect.~\ref{sec:nsiformula}, we have derived general formulas for
the oscillation probabilities including the non-standard effects in
the antineutrino production and detection processes, arguing that
the matter effects throughout the antineutrino propagation do not
play any significant role in short baseline reactor neutrino experiments such as Daya Bay.

In Sect.~\ref{sec:settings} we specified the setting of our main interest corresponding to three different configurations of the NSI parameters. In Sect.~\ref{sec:results} we
performed an illustrative numerical analysis of these settings based on an empirical model of
the reactor antineutrino spectrum at Daya Bay assuming for simplicity that the NSI effects are flavour blind. Taking into account
the statistical uncertainties corresponding to three years of running,  we
have studied how the NSI's could modify the antineutrino energy
spectra and the measured values of the neutrino mixing parameters in
practice. We observe that, under certain
conditions, the Daya Bay experiment can provide hints of such
non-standard effects at more than $90\%$ C.L. {The leading-order systematics has been discussed in brief and it has been shown that it does not play any decisive role in the expected Daya Bay new-physics sensitivity.}

We should also stress the important complementary role the long
baseline experiments, such as, e.g., accelerator experiments, superbeams, beta-beams or a neutrino
factory, could play. Namely, if  $\theta_{13}$ or
$\Delta m^2_{32}$ as determined by Daya Bay differ
significantly from the other results, one would have to take the NSI
effects seriously as one of the possible sources of such a
discrepancy. In that case, a combined analysis of the Daya Bay
data together with the data from the other experiments, including
the NSI's of the kind considered in
this study, would be of utmost importance.

\begin{acknowledgments}
The work is supported by projects ME08076 and MSM0021620859 of Ministry of Education, Youth and Sports of the Czech Republic (R.L. and B.R.), by the Marie Curie Intra European Fellowship
within the 7th European Community Framework Programme
FP7-PEOPLE-2009-IEF, contract number PIEF-GA-2009-253119, by the EU
Network grant UNILHC PITN-GA-2009-237920, by the Spanish MICINN
grants FPA2008-00319/FPA and MULTIDARK CAD2009-00064
(Con-solider-Ingenio 2010 Programme) and by the Generalitat
Valenciana grant Prometeo/2009/091 (M.M.), as well as the ERC under
the Starting Grant MANITOP and the Deutsche Forschungsgemeinschaft
in the Transregio 27 ``Neutrinos and beyond -- weakly interacting
particles in physics, astrophysics and cosmology'' (H.Z.). R.L. and
B.R. acknowledge the hospitality of the Elementary particle theory group of the
 Theoretical physics department of KTH Stockholm during the initial stage of the
project. H.Z. and M.M. are grateful to the Institute for Particle
and Nuclear Physics of the Faculty of Mathematics and Physics at the
Charles University in Prague for the warm hospitality during their
visits. We are indebted to Dmitry V.~Naumov and Marin Hirsch for their insightful comments.  
\end{acknowledgments}

\end{document}